\documentclass{aa}
\usepackage{graphicx}
\usepackage[]{hyperref}
\hypersetup{colorlinks=true,citecolor=blue}
\usepackage{lipsum, babel}
\usepackage{booktabs}
\usepackage{comment}
\usepackage{txfonts}
\usepackage{enumitem} % for itemize label

\usepackage{caption}
\usepackage{subcaption}

\usepackage[version=4]{mhchem}
\usepackage[dvipsnames]{xcolor}
\usepackage{colortbl}

\usepackage{adjustbox}

\usepackage{threeparttable}

\def\coo{CO$_2$\xspace}

\def\od{O($^{1}$D)\xspace}

\def\hh{H$_2$\xspace}

\def\hho{H$_2$O\xspace}
\def\chhhh{CH$_4$\xspace}
\def\chhh{CH$_3$\xspace}

\def\hhco{H$_2$CO\xspace}

\def\chhhoh{CH$_3$OH\xspace}

\def\cchh{C$_2$H$_2$\xspace}
\def\cchhh{C$_2$H$_3$\xspace}
\def\cchhhh{C$_2$H$_4$\xspace}
\def\cchhhhhh{C$_2$H$_6$\xspace}
\def\ccchhhh{C$_3$H$_4$\xspace}
\def\propyne{CH$_3$C$_2$H\xspace}
\def\propdiene{CH$_2$C$_2$H$_2$\xspace}
\def\ccchhhhhh{C$_3$H$_6$\xspace}
\def\ccchhhhhhhh{C$_3$H$_8$\xspace}

\def\chhhcho{CH$_3$CHO\xspace}
\def\nn{N$_2$\xspace}

\def\nhhh{NH$_3$\xspace}
\def\op{O($^3$P)\xspace}
\def\od{O($^1$D)\xspace}
\def\hhs{H$_2$S\xspace}

\def\cme{cm$^{-1}$}

\usepackage{ulem}

\begin{document} 

  \title{Understanding the chemistry of temperate exoplanet atmospheres through experimental and numerical simulations}
  
\titlerunning{Chemistry of temperate exoplanet atmospheres}

  \author{O. Sohier\inst{1}, A. Y. Jaziri\inst{1}, L. Vettier\inst{1}, A. Chatain\inst{1}, T. Drant\inst{2} \and N. Carrasco\inst{1,3}}
     
\authorrunning{Sohier, Jaziri et al.}

  \institute{$^1$ LATMOS/IPSL, UVSQ Universit\'{e} Paris-Saclay, Sorbonne Universit\'{e}, CNRS, Guyancourt, France \\
  $^2$ ETH University, Center for Origin and Prevalence of Life, Department of Earth and Planetary Sciences, 8092 Zurich, Switzerland \\
  $^3$ ENS-Paris-Saclay, Gif sur Yvette, France\\}

  \date{Received 03 October 2025; accepted 17 December 2025}

% \abstract{}{}{}{}{} 
% 5 {} token are mandatory
 
 \abstract{ 

 \textit{Context:}
Characterizing temperate exoplanet atmospheres remains challenging due to their small size and low temperatures. Recent JWST observations provide valuable data, but their interpretation has led to diverging conclusions, highlighting the limitations of observations alone. Complementary approaches combining laboratory experiments and photochemical modeling are essential for constraining atmospheric chemistry and interpreting observations. 

 \textit{Aims:}
This study investigates out-of-equilibrium chemistry in the upper atmospheres of \hh-dominated temperate sub-Neptunes enriched in carbon-bearing species (\chhhh, CO, or \coo). We aim to identify chemical pathways governing the formation and evolution of neutral species and to assess their sensitivity to key parameters such as C/O ratio and metallicity.

 \textit{Methods:}
Our approach combines experimental and numerical simulations on \hh-rich gas mixtures representative of sub-Neptune atmospheres and spanning a wide range of \chhhh, CO, and \coo mixing ratios. We used a cold plasma reactor to simulate out-of-equilibrium upper-atmospheric chemistry. Chemical evolution was tracked by mass spectrometry and infrared spectroscopy. A 0D photochemical model was used to reproduce reactor conditions, guiding interpretation of the key pathways and abundance trends.

 \textit{Results:}
We observed the formation of both reduced and oxidized organic compounds. In \chhhh-rich mixtures, hydrocarbons formed efficiently through methane chemistry, correlating with CH$_4$ concentration and agreeing with models. In more oxidizing environments, particularly \coo-rich mixtures, hydrocarbon formation was inhibited by complex reaction networks and oxidative losses. We find that oxygen incorporation enhances chemical diversity and promotes the formation of oxidized organic compounds of prebiotic interest (\hhco, \chhhoh, \chhhcho), especially in atmospheres containing both \chhhh and \coo. Atmospheres containing \chhhh and CO — which balance carbon and oxygen supply without excessive oxidative destruction — favor efficient production of hydrocarbons and oxidized compounds.

 \textit{Conclusions:}
Out-of-equilibrium chemistry plays a key role in the diversification and organic complexification of temperate exoplanet atmospheres. Combining laboratory experiments with photochemical modeling elucidates pathways to hydrocarbon and oxidized organic formation. Studying detectability of these photoproducts with JWST and new high-resolution ground-based instruments is an important focus for future studies.
}

% % context heading (optional)
% % {} leave it empty if necessary 
%  {}
% % aims heading (mandatory)
%  {}
% % methods heading (mandatory)
%  {}
% % results heading (mandatory)
%  {}
% % conclusions heading (optional), leave it empty if necessary 
%  {Abstract (or use previous aims, methods, etc..)}


  \maketitle

  \nolinenumbers
       
       %
%-------------------------------------------------------------------

\section{Introduction}
Characterizing the atmospheres of temperate exoplanets presents significant challenges due to their small size and relatively cold temperatures, making them difficult to detect and study. These exoplanets, which have an equilibrium temperature below $\sim$500 K, can be classified into three categories: terrestrial planets, super-Earths, and sub-Neptunes. To date, no terrestrial exoplanets, including those in the TRAPPIST-1 system, have been clearly and robustly detected as having an atmosphere (\cite{Gillon_2025} for TRAPPIST-1b and c, \cite{Piaulet_2025} for TRAPPIST-1d, \cite{Espinoza_2025,Glidden_2025} for TRAPPIST-1e). On the other hand, sub-Neptunes and super-Earths — which to this day are the most commonly detected types and therefore dominate the population of known exoplanets \citep{Fressin_2013,Luger_2015,Owen_2016,VanEylen_2018,Hardegree_2019} — have emerged as ideal targets for atmospheric characterization. In particular, those orbiting M dwarf stars have a unique advantage known as the “M dwarf opportunity,” because the smaller size and lower brightness of the host star improve the signal-to-noise ratio during observations. This results in larger transit depths, making it easier to detect and study planetary atmospheres \citep{Cabot_2024,Doyon_2024}. These planets are also particularly intriguing because they have no direct analogs in our Solar System. This raises fundamental questions about the processes that shape their atmospheres, interiors, formation, and habitability \citep{Cabot_2024}. In addition, a significant number of planets orbiting M dwarfs reside in their star’s temperate zone \citep{Dressing_2015}, making the question of their habitability all the more interesting. This habitability is challenged by the unique environmental conditions around M dwarfs, including intense stellar flares, high-energy particle events, and limited ultraviolet flux to drive prebiotic chemistry \citep{Scalo_2007,Ricker_2015,Shields_2016,Rimmer_2018}.

Recent advancements in observational techniques, particularly with instruments on the James Webb Space Telescope (JWST), have made these planets favorable targets. In particular colder exoplanets, with temperatures between 500 and 1000 K, tend to have clearer atmospheres, as they are less obscured by clouds and hazes \citep{Morley_2015,Brande_2024,Yang_2024,Gao_2021}. The advancements have already led to the successful characterization of two temperate sub-Neptune exoplanetary atmospheres, marking significant progress in this field. The analysis of transit spectroscopy data from the atmospheres of K2-18 b \citep{Madhusudhan_2023} and TOI-270 d \citep{Benneke_2024,Holmberg_2024} has highlighted atmospheric compositions dominated by hydrogen and containing a few percent of carbon-bearing species, such as methane or carbon dioxide. Despite these revolutionary new data, their atmospheric composition is still poorly constrained and highly debated. 

In the case of K2-18 b, initial JWST observations in the 0.9–5.2 µm range combined with data from NIRISS SOSS and NIRSpec G235H instruments acquired in 2023 suggest the presence of \chhhh and \coo at comparable levels, around 1\% each \citep{Madhusudhan_2023}. Subsequent studies incorporating non-equilibrium chemistry have led to a revised interpretation, indicating a higher level of \chhhh and a lower abundance of \coo \citep{Wogan_2024}. This reinterpretation is supported by a new analysis of the JWST data, which has called into question the initial detection of \coo as a whole \citep{Schmidt_2025}. 
In 2024 and early 2025, four new transit observations were acquired with the NIRSpec G395H instrument and presented by \cite{Hu_2025}. By stacking all NIRSpec spectral data obtained to date, they report a significant detection of both \chhhh and \coo, with upper limits of approximately 10\% and 1\% respectively, supporting the hypothesis of a predominance of methane. 
Meanwhile, initial data were also acquired with the JWST MIRI LRS instrument in the 6–12 µm range, and primary analysis of the data did not provide strong evidence for \chhhh or \coo \citep{Madhusudhan_2025}, which do not show clear bands in this range.

The detection of signatures of minor atmospheric constituents also remains largely unconstrained by these low-resolution observations. Non-equilibrium chemistry models help characterize the complete atmospheric composition, but they require observational constraints such as metallicities and C/O ratios, which are also poorly known. They predict the abundance of various components, even minor ones for which signatures remain undetectable in observations. According to these simulations, \chhhh, \coo, CO, \nn, \nhhh, \hho, \cchhhh, and \cchhhhhh could be major species, up to the percent range \citep{Jaziri_2025,Luque_2025}. \cite{Jaziri_2025} also predict that oxidized species such as \hhco, \chhhoh, and \chhhcho can reach mixing ratios close to the parts-per-million level. These molecules are of significant prebiotic interest. \hhco is important as a possible precursor to sugars through the formose reaction, which is key for the origin of the RNA or pre-RNA world \citep{Butlerow_1861,Cleaves_2008}. \chhhoh is a key synthon in the RNA world \citep{Mathew_2022}.  Aldehydes, such as \hhco and \chhhcho, are key precursors of several amino acids through the Strecker synthesis \citep{Strecker_1850}. The atmospheric photochemical production of these oxidized compounds has already been identified as one of the main sources of these compounds in the atmosphere of primitive Earth  \citep{Pinto_1980,Junshan_Pinto_1989}. It is therefore worthwhile to study their formation pathways in temperate exoplanetary atmospheres.

If we are to overcome the major scientific challenge of characterizing the chemistry of temperate exoplanet atmospheres, a synergy between the observational, modeling, and experimental communities is required. Initiatives of this type are already underway in the community. One example is the attempt to obtain experimental data on the temperature-dependent photo-absorption cross sections of key molecules such as \coo, which provide essential data for atmospheric models of hot exoplanets \citep{Venot_2013,Venot_2018}. For the cooler exoplanets, studies have been carried out on the detectability of ions in the thermosphere of sub-Neptunes by combining laboratory experiments, chemical modeling, and simulated observations as well as in the atmospheres of super-Earths by combining laboratory measurements and a 0D photochemical model \citep{Bourgalais_2020,Bourgalais_2021}.

A few experiments have been developed for the characterization of the atmospheres of exoplanets, and they mainly focus on the formation and properties of haze and aerosols in the atmospheres of super-Earths and sub-Neptunes \citep{Gavilan_2018,He_2018,He_Horst_2018_lab,Horst_2018,Moran_2020,Escobar_2025} or hot Jupiters \citep{Fleury_2019,Fleury_2020,Fleury_2023}. Studies focusing on the neutral gas phase chemistry in the atmospheres of cool exoplanets are quite rare. \cite{He_Horst_2019_gas} conducted laboratory experiments on a wide range of exoplanet atmospheres, with a wide variety of metallicities (100×, 1000×, and 10000× solar metallicity) and temperatures (at 300, 400, and 600 K) and for warmer \hh-rich atmospheres (800 K, 100x, and 1000x solar metallicities \citep{He_2020}). They explored the effect of different types of irradiation, using a cold plasma or a UV photon lamp as energy sources. They found that a complex non-equilibrium chemistry occurred, leading to the formation of potential haze-forming precursors. \cite{Wang_2025} explored \coo-rich atmospheres at 300 and 500 K, with a 2000x solar metallicity. Under plasma irradiation, they observed the formation of various hydrocarbons and oxygen- and nitrogen-containing species, and they identified reactive gas precursors for haze formation such as \cchhhh, \hhco, and HCN. However, the analytical diagnosis of all these studies was based solely on residual gas analyzer mass spectrometry (RGA MS), which led to an uncertain identification of some compounds. \cite{Reed_2024} studied the UV irradiation of \hhs/\chhhh/\nn gas mixtures, with or without \coo, at an ambient temperature (around 300 K). They analyzed the gas-phase products using gas chromatography equipped with sulfur chemiluminescence detection (GC-SCD), a diagnostic tool less ambiguous than RGA MS. They were thus able to demonstrate the abiotic production of several organosulfur gases, limiting their robustness as biosignatures.

The aim of this work is to propose chemical mechanisms occurring in a wide range of possible atmospheric compositions of temperate sub-Neptunes. In this context, we carried out parallel experimental and 0D numerical simulations to understand the complex out-of-equilibrium chemistry occurring in the upper layers of the \hh-dominated atmospheres of temperate exoplanets. These low-pressure atmospheric layers correspond to those probed by JWST observations and were exposed to energetic radiation that triggers photochemical processes. As the atmospheric composition of these exoplanets is unconstrained, we carried out a study on various \hh/\chhhh/CO/\coo gas mixtures. The C/O ratio (from 0.5 to infinity) and metallicity (from 5 to 50 times the solar metallicity) of these mixtures are variable. The C/O ratio measures the relative abundance of carbon compared to oxygen, while metallicity measures the abundance of elements heavier than hydrogen and helium. To vary these parameters in our \hh-dominated gas mixtures, we used \chhhh, CO, or \coo as the main component as well as different combinations of these carbon compounds. We chose to focus on these reactants because they allow us to obtain a wide range of possible atmospheres based on the observational data currently available for K2-18 b. Of course, the reactants are relatively simplistic atmospheric analogs that do not take into account other elements such as nitrogen or sulfur, which may also be present in this type of atmosphere. Other oxygen-bearing molecules, such as water, could also be present in significant quantities in this atmosphere, although the upper layers may instead be dried out by cold trap processes \citep{Charnay_2021}. Nevertheless, these atmospheric analogs already allow us to cover a wide range of atmospheric possibilities and to identify a great molecular diversity. A key aspect of this work lies in the combination of experiments and numerical simulations, which enables us to extract abundance trends and chemical formation pathways. Identifying key mechanisms enables us to understand the role of different carbon-bearing reactants (\chhhh, CO, \coo) in organic growth processes (i.e., carbon chains elongation) and to study the effect of oxygen incorporation on chemical diversification.

\section{Methods}
\label{sec\string:expsetup}
We studied the chemical evolution of atmospheric analog mixtures when irradiated in a cold plasma setup, mimicking photochemistry in the upper atmosphere. Complementary diagnoses were performed using mass spectrometry (MS) and infrared spectroscopy. In parallel, a 0D photochemical model predicted abundance values and reaction pathways, which we systematically compared with experimental results.

\subsection{Mixtures of atmospheric analogs}
Different gas mixtures were chosen to study a wide range of potential atmospheres, based on information gathered from observations of K2-18 b. These mixtures are dominated by hydrogen, with carbon-bearing compounds ranging from highly reduced carbon (\chhhh) to an intermediate state (CO) to fully oxidized carbon (\coo). To enable comparisons with planetary atmospheric models, the C/O ratios and metallicities of our mixtures are provided for reference. It should be noted that our compositions are set by the continuous gas flows in an open system. The gas residence time, i.e., the time spent by the gas in the irradiated medium, is set by the pressure and gas flow in our experiments and thus differ from the residence time found in atmospheric models. Thus, the C/O ratios given are enriched in \chhhh, CO and \coo, compared to atmospheric models. All mixtures are summarized in Table \ref{tab:compositions}.

The most reduced atmospheres are composed of \hh and \chhhh. The relative proportions of \hh and \chhhh vary in the experiments (see Table \ref{tab:compositions} for the relative mixing ratios). We refer to these mixtures as reduced mixtures. This type of atmosphere helps us characterize organic growth in the most reducing environment.

The more oxidized atmospheres are composed of \hh and CO or \coo. The relative mixing ratios are shown in Table \ref{tab:compositions}. We refer to these mixtures as oxidized mixtures. This type of atmosphere allows us to explore the balance between adding carbon and oxygen atoms to the system. 

The composite atmospheres combine the conditions of the reduced and oxidized atmospheres described above. They are composed of \hh, \chhhh, CO and/or \coo. A wide range of oxidation states is studied, as shown in Table \ref{tab:compositions}. 

\begin{table*}[h!]
  \caption{Experimental compositions in percentages.}
  \centering
  \resizebox{\textwidth}{!}{% 
  \begin{tabular}{l|cccc|cccc|cccc|ccccc|ccccc|cc|}
    %\multicolumn{20}{c}{GJ~1214~b}\\ [0.3cm] \cline{3-20}
    \cline{2-25}
    \rule{0pt}{4ex} & \multicolumn{4}{c|}{\textbf{\textit{Reduced mixtures}}}  & \multicolumn{8}{c|}{\textbf{\textit{Oxidized mixtures}}}  & \multicolumn{12}{c|}{\textbf{\textit{\chhhh/CO/\coo mixtures}}} \\ \cline{2-25}
    \hline
    \multicolumn{1}{|l|}{\hh} & 99 & 97 & 95 & 90 & 99 & 97 & 95 & 90 & 99 & 97 & 95 & 90 & 98 & 94.5 & 94 & 94 & 90 & 98 & 94.5 & 94 & 94 & 90 & 97,5 & 89,5 \\
    \hline
    \multicolumn{1}{|l|}{\chhhh} & 1 & 3 & 5 & 10 & \cellcolor{Gray} & \cellcolor{Gray} & \cellcolor{Gray} & \cellcolor{Gray} & \cellcolor{Gray} & \cellcolor{Gray} & \cellcolor{Gray} & \cellcolor{Gray} & 1 & 5 & 1 & 5 & 5 & 1 & 5 & 1 & 5 & 5 & 1 & 5 \\
    \hline
    \multicolumn{1}{|l|}{CO} & \cellcolor{Gray} & \cellcolor{Gray} & \cellcolor{Gray} & \cellcolor{Gray} & 1 & 3 & 5 & 10 & \cellcolor{Gray} & \cellcolor{Gray} & \cellcolor{Gray} & \cellcolor{Gray} & 1 & 0.5 & 5 & 1 & 5 & \cellcolor{Gray} & \cellcolor{Gray} & \cellcolor{Gray} & \cellcolor{Gray} & \cellcolor{Gray} & 1 & 5 \\
    \hline
    \multicolumn{1}{|l|}{\coo} & \cellcolor{Gray} & \cellcolor{Gray} & \cellcolor{Gray} & \cellcolor{Gray} & \cellcolor{Gray} & \cellcolor{Gray} & \cellcolor{Gray} & \cellcolor{Gray} & 1 & 3 & 5 & 10 & \cellcolor{Gray} & \cellcolor{Gray} & \cellcolor{Gray} & \cellcolor{Gray} & \cellcolor{Gray} & 1 & 0.5 & 5 & 1 & 5 & 0.5 & 0.5 \\
    \hline
    \multicolumn{1}{|l|}{C/O ratio} & $\infty$ & $\infty$ & $\infty$ & $\infty$ & 1 & 1 & 1 & 1 & 0.5 & 0.5 & 0.5 & 0.5 & 2	& 11 & 1.2 & 6 & 2 & 1 & 5.5 & 0.6 & 3 & 1 & 1.25 & 1.75 \\
    \hline
    \multicolumn{1}{|l|}{Metallicity (compared to solar)} & 5.6 & 14.8 & 22.1 & 35.1 & 9.2 & 22.6 & 31.7 & 45.4 & 13.6 & 30.2 & 40.0 & 53.0 & 13.7 & 24.8 & 33.8 & 27.2 & 41.0 & 17.5 & 26.2 & 41.5 & 29.7 & 46.6 & 18.9 & 42.9 \\
    \hline
  \end{tabular}
  }
  \label{tab:compositions}
\end{table*}

\subsection{Atmospheric simulation chamber}

Experimental simulations of the upper atmospheric layers of temperate exoplanets were carried out using an experimental cold plasma device, the PAMPRE reactor (for Production d'Aérosols en Microgravité par Plasma REactif \citep{Szopa_2006}). This setup was initially developed to simulate the atmospheric chemistry of Titan, and the production of aerosol analogs (\textit{tholins}) created by photochemistry in this atmosphere \citep{Szopa_2006,Quirico_2008,Sciamma_2010,Carrasco_2012,Mahjoub_2012,Fleury_2014,Dubois_2019,Dubois_2020}. The use of this experimental platform has been extended over the years to include other atmospheres, such as early Earth \citep{Fleury_2015,Fleury_2017} or Pluto \citep{Jovanovic_2021}. More generally, the versatility of the experimental setup means we can study any temperate atmospheric chemical composition, which make it suitable for studying exoplanets. Initial research has already been carried out on organic aerosols produced in the atmospheres of Earth-like exoplanets \citep{Gavilan_2017,Gavilan_2018}. 

PAMPRE induces a cold capacitively coupled radiofrequency plasma in a gaseous mixture, outside of thermodynamic equilibrium. This reproduces the energetic processes occurring in the upper layers of a temperate atmosphere exposed to energetic stellar particles and UV irradiation. It has been previously shown that the energy distribution of electrons in a nitrogen-dominated plasma reproduces the energy distribution of solar photons fairly accurately, with a slightly higher contribution in the VUV \citep{Szopa_2006}. The stellar spectra of M dwarfs — around which the temperate exoplanets that interest us orbit — have higher UV radiation contributions than the Sun \citep{France_2013}. This experimental setup is therefore appropriate for simulating the energetic environment of these exoplanetary atmospheres. Although our plasma is dominated by hydrogen, which may slightly alter the energy distribution of electrons, it remains adequate. These energetic processes trigger an imbalance in chemistry, dissociating gas molecules and allowing fragments to reorganize into new species. In specific cases, products even aggregate into solid particles, the tholins.

\begin{figure} [ht!]
  \centering
  \includegraphics[width=0.5\textwidth]{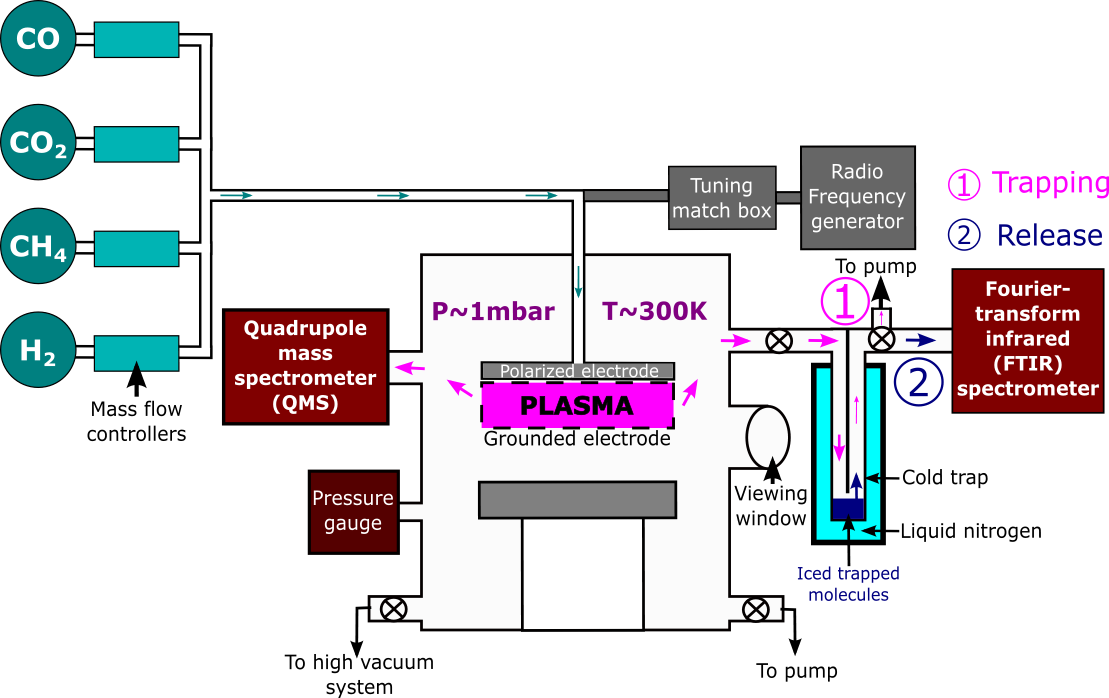}
  \caption{Scheme of the plasma reactor PAMPRE.}
  \label{pampre}
\end{figure}

The setup consists of a cylindrical stainless steel reactor, 40 cm high and 30 cm in diameter, connected to a quadrupole mass spectrometer (QMS). High-pressure gas cylinders are connected to the setup via an injection manifold, enabling gas mixtures to be introduced into the main chamber. MKS mass flow meters, rated at 5, 10 and 100 standard cubic centimeters per minute (sccm) with a full-scale accuracy of 1\%, control the injection rates. The plasma is confined in a stainless steel cage measuring 13.8 cm in diameter and 5 cm in height, pierced with small holes to allow the gas to pass through. The top of this cage is the cathode, measuring 12.6 cm in diameter, polarized and powered by a 13.56 MHz generator. The gas mixture is injected into the reactor from the top of the chamber through the cathode, which is equipped with three superimposed grids to ensure a continuous, homogeneous injection flow. Gas flow rate is set at 40 sccm. The central grid, with a mesh size in the order of the Debye length ($\sim$0.1 mm), prevents the plasma from flowing back into the gas injection system. The gas mixture is extracted from the bottom of the chamber by a primary vane pump (Adixen by Pfeiffer Vacuum Pascal 2015 SD). This enables an open continuous-flow system, at 0.98 mbar and at ambient temperature ($\sim$300 K). A window allows visual monitoring of the plasma. Plasma discharge is only used in continuous mode in these experiments, with a power of 30 W. Between each experiment, the reactor is pumped with a turbomolecular pump (Adixen ATP 80) to achieve a secondary vacuum of around 1.5×10$^{-6}$ mbar. A schematic diagram of the experimental setup is shown in Figure \ref{pampre}.

\subsection{Mass spectrometry} 

Mass spectrometry is used as an analytical method to determine the molecular species produced. PAMPRE is equipped with a Hiden EQP system, which is used in residual gas analysis (RGA) mode, since only neutral species are considered in this study. A collector head is used to sample gaseous species through a 100 µm diameter orifice. The pressure inside the mass spectrometer is kept below 5x10$^{-6}$ mbar. To study neutral species, an internal ionizer is used, with electron energy set at 70 eV. The resulting ions are then filtered and focused before entering the triple-filter mass analyzer, which detects masses from 1 to 200 m/z (mass-to-charge ratio) with a resolution of 1 atomic mass unit (amu). Finally, the ions arrive at an ion-counting detector — a secondary electron multiplier.

Several molecular fragments at different m/z values are generated and form a "fingerprint" specific to each molecule and incident electron energy, known as the fragmentation pattern. A full mass spectrum thus corresponds to the superposition of the fragmentation patterns of all the species present in the analyzed gas mixture. Reference fragmentation patterns were measured directly in the lab for the reactive compounds, i.e., \hh, \chhhh, CO and \coo. The reference fragmentation patterns of the products are extracted from the NIST Chemistry WebBook\footnote{\url{https://webbook.nist.gov/chemistry/}} \citep{NIST}.

All m/z over a selected mass range are scanned in BAR mode (Batched Analysis of Repetitions). This produces a histogram of peak intensities. Each mass spectrum contains 20 accumulated scans, over a mass range from 1 to 100 amu. A mass spectrum is taken at the initial state (spectrum OFF), when the gaseous mixture is injected inside the reactive chamber at 0.98 mbar, while the plasma discharge is still switched off. The plasma is switched on and after five minutes, a spectrum is taken (spectrum ON). To ensure that we had reached a stable steady state, we performed multiple ion detection when we activated the plasma. This allowed us to track several m/z ratios over time, thereby monitoring consumption and production and determining when we reach a stable state. In all our experiments, we observed stabilization after a maximum of 200 seconds.

\begin{figure}[ht!]
  \centering
  \includegraphics[width=0.49\textwidth,trim = 0cm 0cm 0cm 0cm, clip]{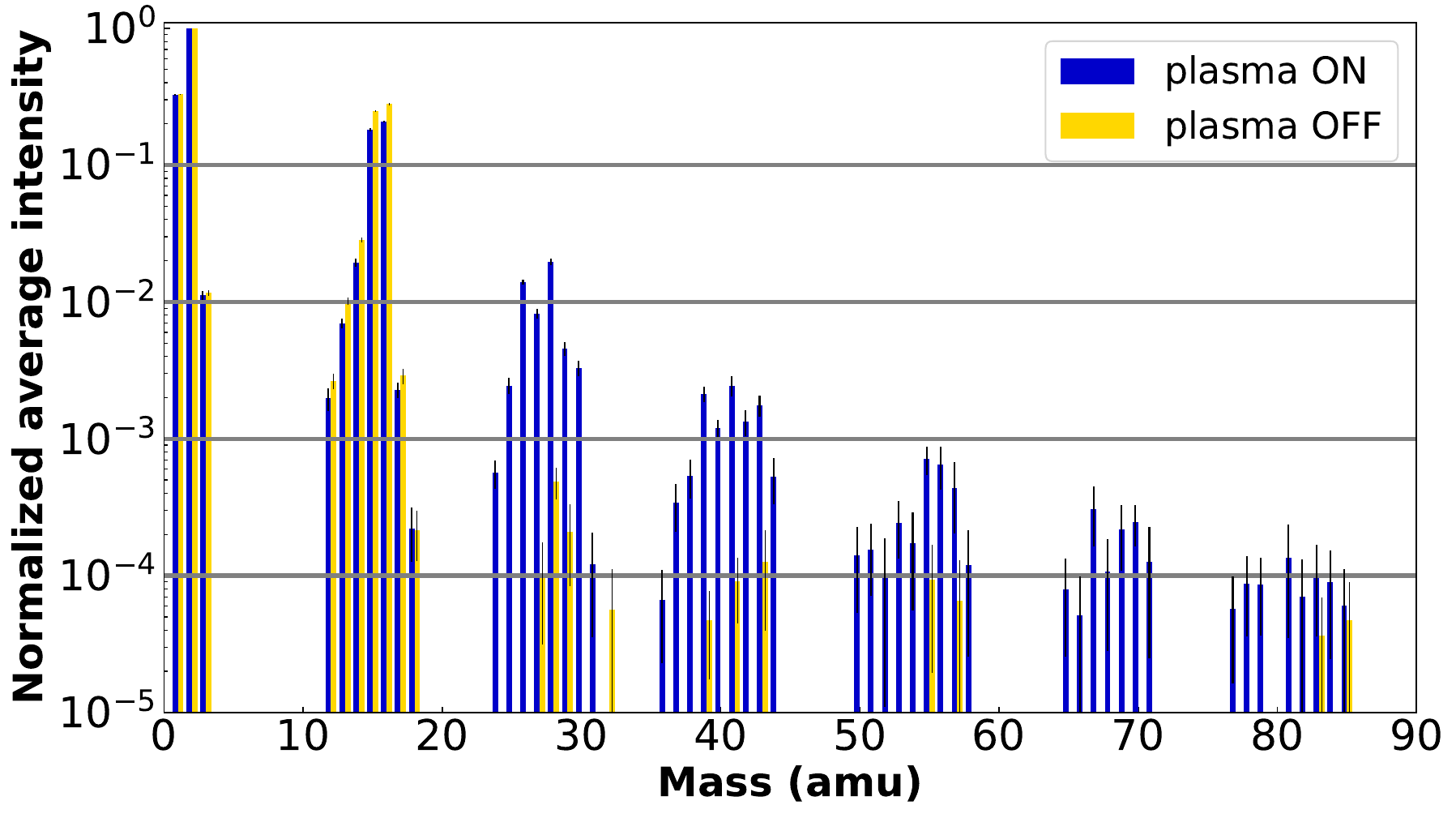}
  \caption{Mass spectrum of a gas mixture of 95\% \hh plus 5\% \chhhh before (gold) and during irradiation (blue).}
  \label{carbon_chains_ms}
\end{figure}

The evolution of the gas composition can be seen by our plotting of the superposition of the two ON and OFF spectra (Figure \ref{carbon_chains_ms}). Our plotting of the difference of intensity between these two spectra (ON-OFF) reveals the production and destruction of the chemical species.

The measured mass spectra are processed first by removing the noise from the data: all intensities below 10 counts/s are removed, as the instrument's sensitivity is 10$^{6}$ counts/s maximum. Each spectrum is then normalized by its maximum intensity. In the direct-flow experiments, the maximum intensity is always at m/z 2 amu, corresponding to H$_2$. In trap-and-release experiments, the maximum intensity depends on the composition. Mean values and standard deviation at 95\% are calculated from the 20 cycles for each m/z. If the difference between the average intensity and the standard deviation is below 10$^{-6}$, it is considered as noise and removed from the data.

While the reconstruction of a mass spectrum with fragmentation patterns works well with a few compounds, it becomes much more difficult to analyze and identify species in more complex mixtures with dozens of products. This is due to the superposition of several fragmentation patterns and the variability associated with these patterns — which depends on the geometry of the ion source, the mass-dependent filtering, and detection efficiency of the instrument. We therefore need a complementary analytical method in order to make unambiguous identifications.

\subsection{Infrared spectroscopy } 
\label{sec:irmethod}

Infrared spectroscopy was used to help identify species in \chhhh/CO/\coo mixtures. The PAMPRE setup was connected to a multipass infrared white cell with an optical beam length of 10 m. The cell was positioned in the sample compartment of an FTIR spectrometer (Nicolet 6700, Thermo Fisher). 

Gas mixtures flow directly through this cell to characterize the dissociation rate of reactants and the formation of major products. However, the concentration of minor products is too low relative to the instrument's thresholds to allow their identification. To overcome this limit, we use a cold trapping strategy to concentrate our products before IR identification. A glass trap installed at the outlet of the system is immersed in liquid nitrogen (78 K) (see Figure \ref{pampre}), ensuring the entire flow passes through it, condensing most of the gases. \hh, \chhhh and CO, with condensation temperatures below 78 K, are not trapped, which helps to concentrate minor products. However, precise quantification is not possible due to this bias-inducing differentiated condensation, but robust product identification is already an important step forward. The irradiated gas stream is allowed to pass through the trap for 20 min; trapping is then stopped, the trap is heated to sublimate the condensed species, the gases are released into the multipass cell, and heated to 70°C so that water vapor does not condense on the mirrors. The pressure is measured inside the cell and it does not exceed a few millibars. Since there is no longer any irradiation, we remain at low pressure and do not reach very high temperatures, so we expect the kinetics to be slow and no significant new reactions to occur. Infrared spectra are measured through the software OMNIC (Thermo Scientific), with a resolution of 0.5 \cme using the Graymann-Harris apodization function \cite{Harris_1978}. 150 scans were accumulated for each measurement, averaged, and combined into a spectrum.

Preprocessing the spectra, using OMNIC software, transforms the single-beam data acquired into transmittance or absorbance spectra, by dividing them by blank spectra previously acquired under high vacuum conditions (less than 10$^{-5}$ mbar) and by applying a manual baseline correction. The IR spectra were compared with those calculated at the experimental pressure (1 mbar) and temperature (340 K), based on the opacities of several species (Figure \ref{IR_ref_all}). The opacities are taken from the databases DACE \footnote{\url{https://dace.unige.ch/opacityDatabase/?}} and HITRAN \citep{Gordon2017_HITRAN}. However, these identifications were limited by the availability of molecules — organic compounds with three or more carbon atoms are not included. For more complex molecules, identification is performed using the NIST database, but this does not allow for temperature and pressure dependence to be taken into account.

\subsection{Atmospheric numerical simulations}

To model the chemistry occurring in the reactor chamber, we used the 0D photochemical model \textit{ReactorUI} \citep{Peng_2014,Pernot_2023}, as described in \cite{Peng_2014}. This complementary approach to experimental characterization helps to identify the species produced and to elucidate the main chemical pathways leading to their formation \citep{Bourgalais_2020}.

The model reproduces experimental conditions, taking into account chamber geometry, gas flows and the distribution of incoming plasma energy, which has been approximated using solar spectra. This is a significant assumption that can influence abundance but still allows pathways to be identified. After adjusting all the parameters, the integration system converges to a steady state.

The chemical network was originally developed for the CHN atmosphere of Titan with neutral-neutral reactions \citep{Hebrard_2007,Hebrard_2009,Dobrijevic_2016}, ions \citep{Carrasco_2008,Plessis_2012} and photo-processes (from the Leiden \citep{Heays_2017} and SWRI \citep{Huebner_2015} databases). For the studies of the formation of more complex oxygenated molecules, such as \hhco and \chhhoh, we extended the chemical network with that of CHO \textit{VULCAN} \citep{Tsai_2021}, specific to sub-Neptunes, based on \cite{Venot_2020a}.

Simulations of each experimental composition were run. Considering chemical rate uncertainties with a Monte Carlo approach, we perform 500 runs for each configuration. This Monte Carlo approach to characterizing uncertainties has already been used in several atmospheric models of Solar System objects, such as Triton (250 runs, \cite{Benne_2022}), Titan (500 runs, \cite{Hebrard_2009}) or Neptune (1000 runs, \cite{Dobrijevic_2010}). As a result, we evaluate the final species abundance values with its uncertainty. These simulations also provide information on the contributions of specific reactions to the formation or consumption of a given molecule, enabling reaction paths to be extracted.

\section{Results}

\subsection{Identification of main compounds}
\label{sec:identification}
Table \ref{tab: features} details the value of the main IR and MS features of the major products. Figure \ref{IR_ref_all} shows the reference IR spectra extracted from the HITRAN database for the molecules containing up to two carbon atoms considered in this study.

\subsubsection{Reduced organic compounds}
We observe the formation of reduced organic compounds, composed only of carbon and hydrogen atoms, forming chains with up to six carbon atoms. In MS, groups of peaks form at m/z up to 80 when a 95\% \hh + 5\% \chhhh gas mixture is irradiated (Figure \ref{carbon_chains_ms}). The specific fragmentation patterns (FP) of reduced compounds with two and three carbon atoms (C2 and C3) are shown in Figure \ref{FP_organics}. These patterns are normalized to the mass spectrum measured in the irradiated gas mixture of 95\% \hh + 5\% \chhhh, and directly compared with the corresponding mass spectrum. In the case of reduced atmospheres, C2 identification using MS alone is possible, particularly for \cchh and \cchhhhhh, which exhibit strong characteristic peaks. Infrared spectroscopy is used to determine the nature of the chemical bonds between carbon atoms, distinguishing between single, double and triple bonds. The IR features used to identify acetylene (\cchh), ethene (\cchhhh), propyne (\ccchhhh) and propene (\ccchhhhhh) are shown in Figure \ref{IR_spectra_reduced}. 

\subsubsection{Oxidized organic compounds}
 
\begin{figure*}[ht!]
  \begin{subfigure}{0.49\textwidth}
  \includegraphics[width=\textwidth]{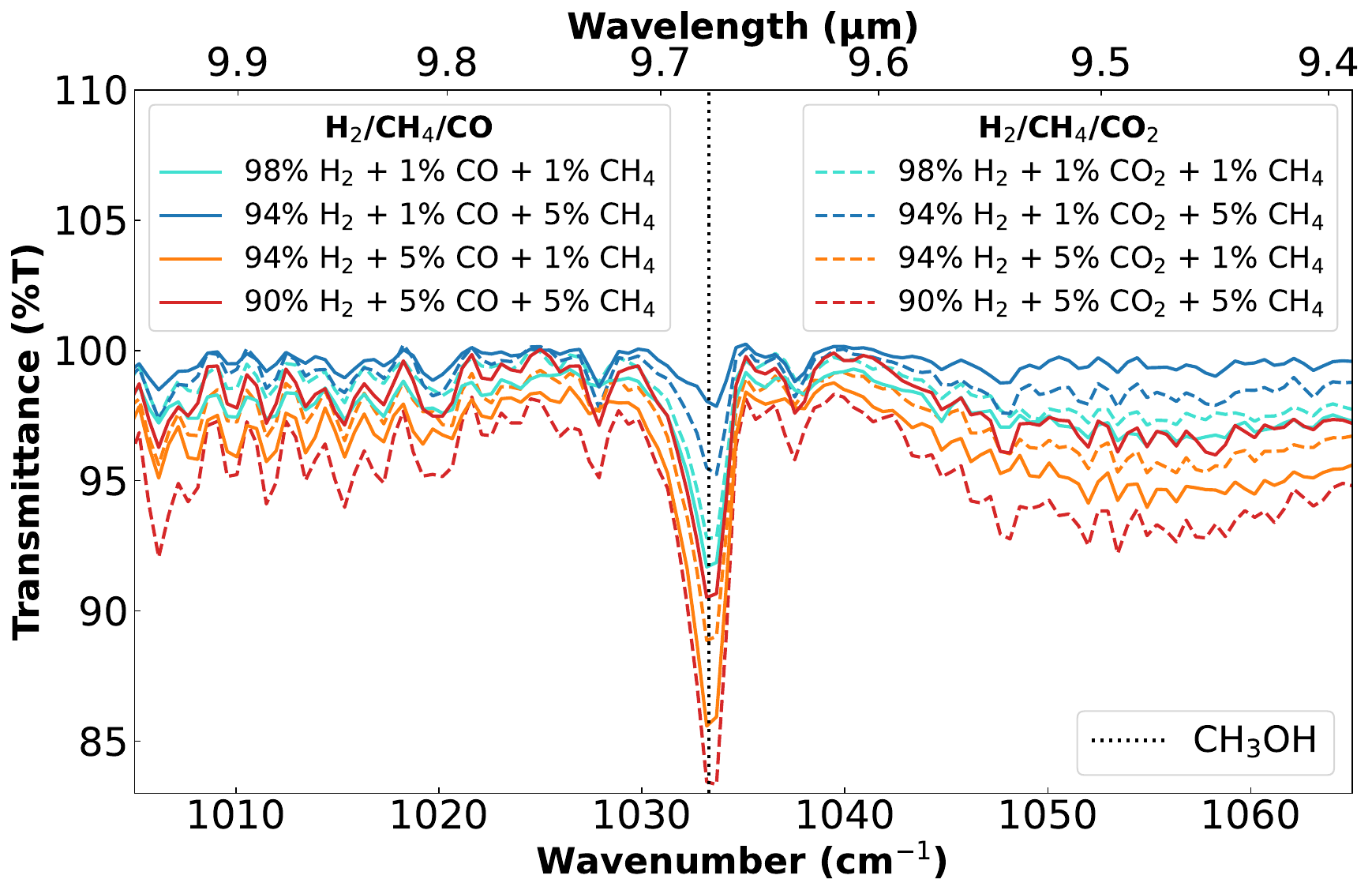}
  \caption{Infrared feature at 1033 \cme (9.68 µm), attributed to \chhhoh.}
  \label{IR_CH3OH}
  \end{subfigure}
  \begin{subfigure}{0.49\textwidth}
  \includegraphics[width=\textwidth]{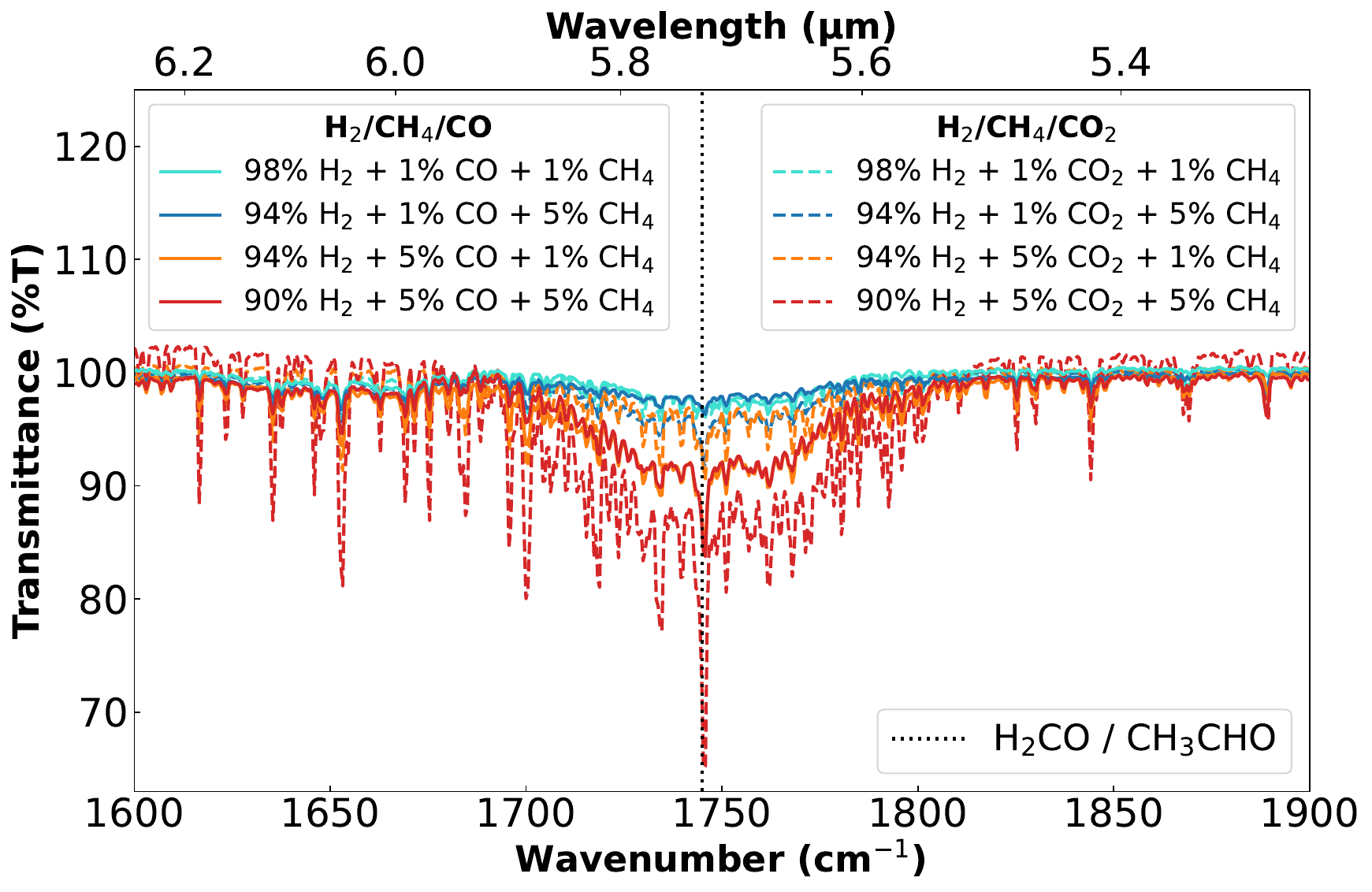}
  \caption{Infrared feature at 1745 \cme (5.73 µm), attributed to aldehydes.}
  \label{IR_ald}
  \end{subfigure}
  \caption{Infrared features of the oxidized organics.}
  \label{IR_spectra_oxidized}
\end{figure*}

Using IR, we could see in CO and \coo-rich mixtures the robust formation of the oxidized organic compounds methanol (\chhhoh), formaldehyde (\hhco) and acetaldehyde (\chhhcho). The features used for their identification are shown in Figure \ref{IR_spectra_oxidized}, at 1033 \cme (9.68 µm) for \chhhoh and at 1745 \cme (5.73 µm) for the aldehydes (\hhco and \chhhcho). In MS, the production of \chhhoh is identified by the appearance of the characteristic peaks at m/z 31 and 32. The fragmentation patterns of \hhco and \chhhcho overlap with those of \cchhhhhh and \ccchhhhhh, which makes their identification only through MS a source of uncertainty.  

\subsubsection{Other oxidized compounds}
Depending on the degree of oxidation of the reactants, the formation of other oxidized molecules is identified. \hho is identified in MS with peaks at m/z 17 and 18, and presents wide features in IR. CO is identified with a specific peak at m/z 28 in MS, but is not trapped for IR measurements as discussed in section \ref{sec:irmethod}. \coo is identified in MS with a peak at m/z 44, and in IR with a narrow peak at 668 \cme (14.97 µm) and a broad one from 2200 to 2400 \cme (4.17-4.55 µm).

\begin{figure*}
  \begin{subfigure}{0.49\textwidth}
  \includegraphics[width=\textwidth]{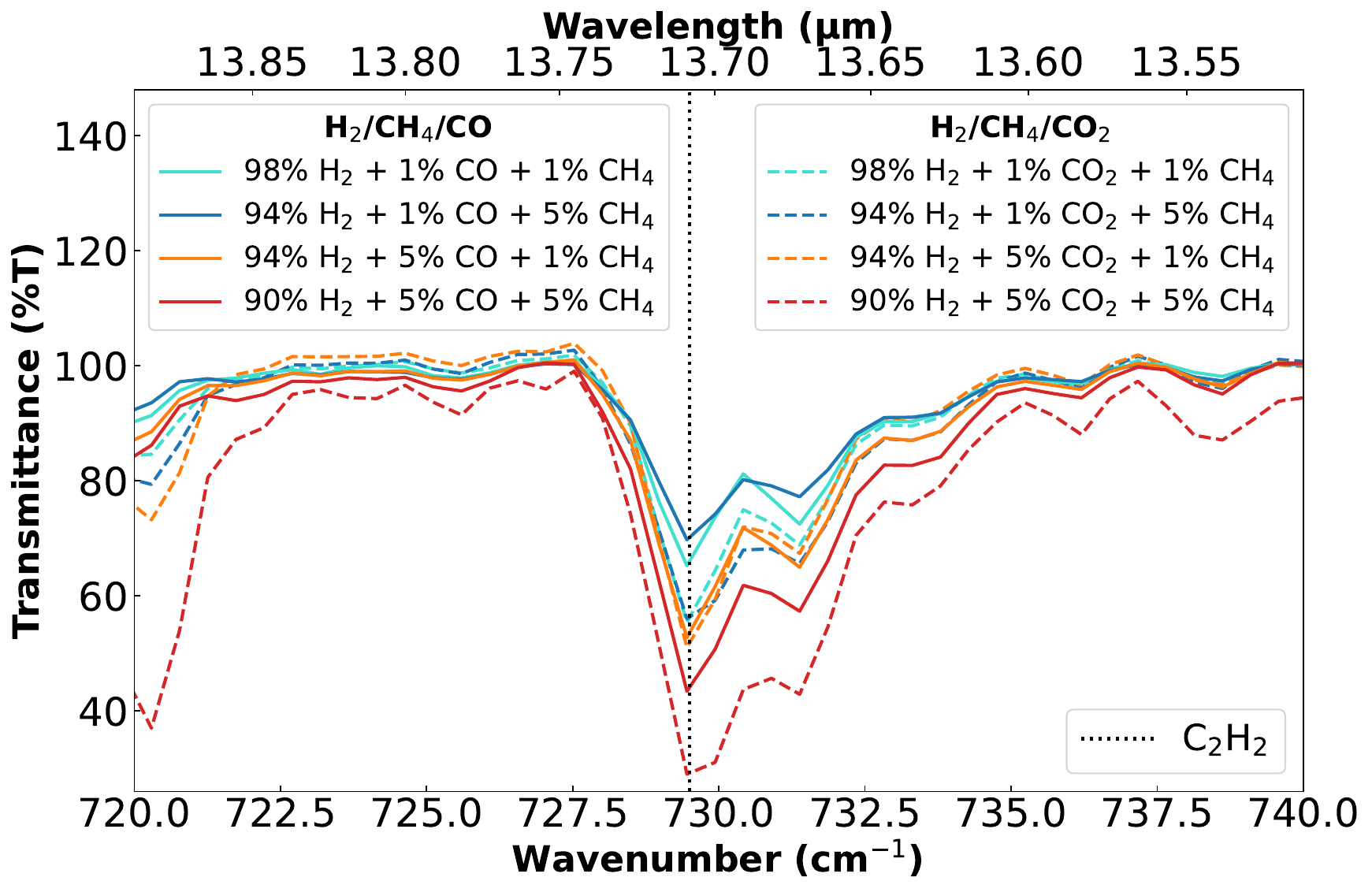}
  \caption{Infrared feature at 730 \cme (13.70 µm), attributed to \cchh.}
  \label{IR_730}
  \end{subfigure}
  \begin{subfigure}{0.49\textwidth}
  \includegraphics[width=\textwidth]{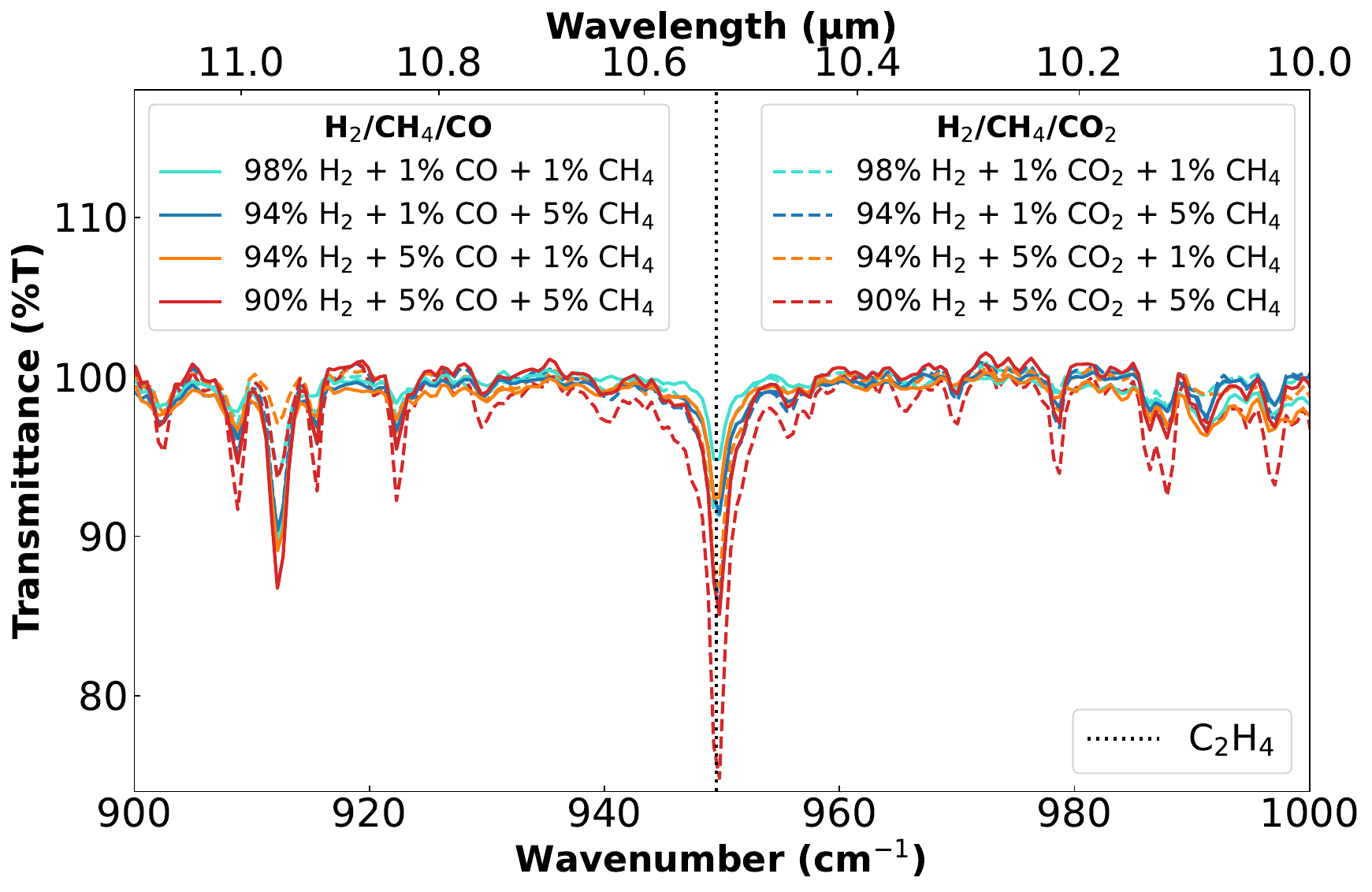}
  \caption{Infrared feature at 950 \cme (10.53 µm), attributed to \cchhhh.}
  \label{IR_950}
  \end{subfigure}
  \begin{subfigure}{0.49\textwidth}
  \includegraphics[width=\textwidth]{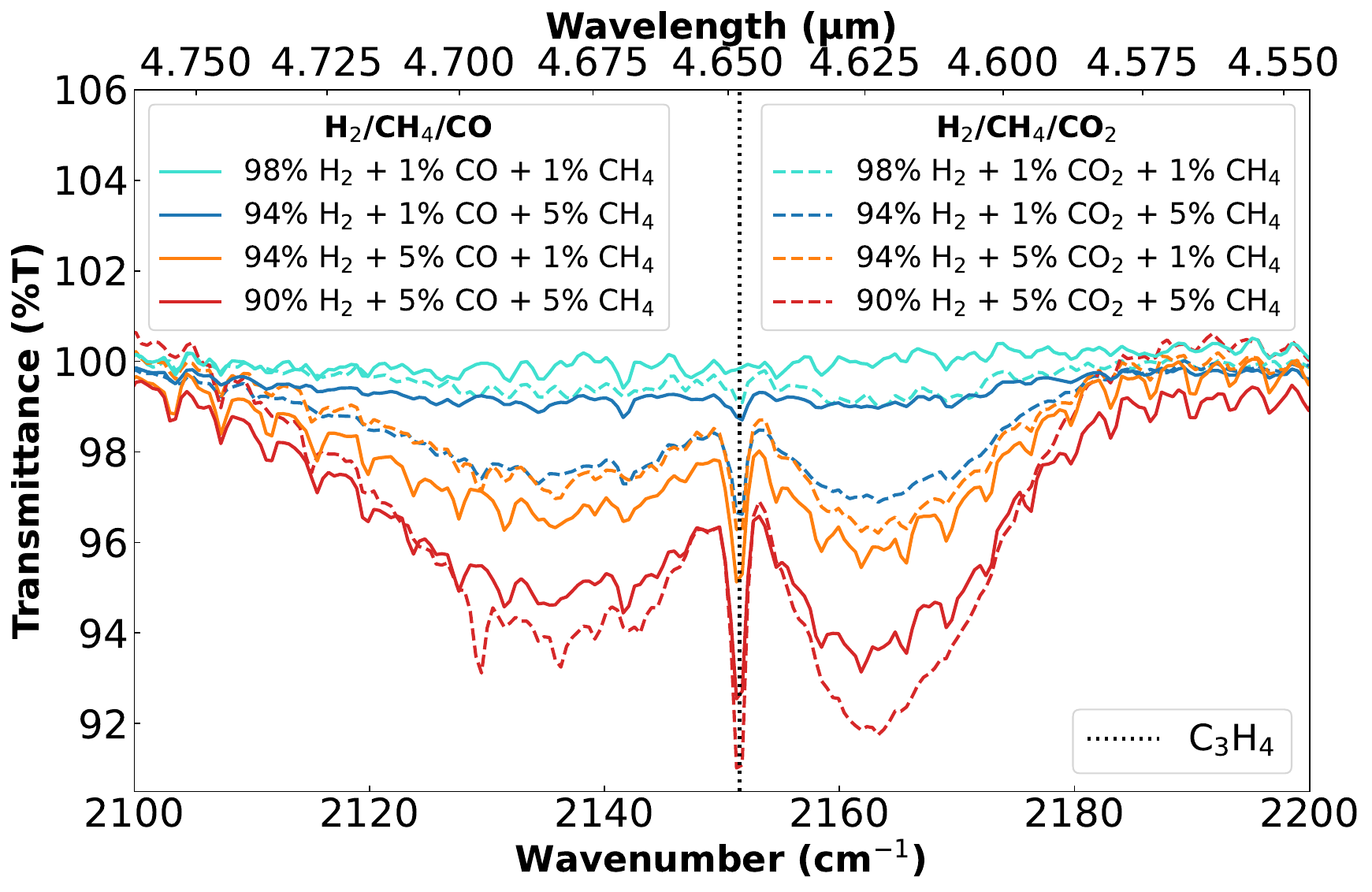}
  \caption{Infrared feature at 2150 \cme (4.65 µm), attributed to \ccchhhh.}
  \label{IR_2150}
  \end{subfigure}
  \begin{subfigure}{0.49\textwidth}
  \includegraphics[width=\textwidth]{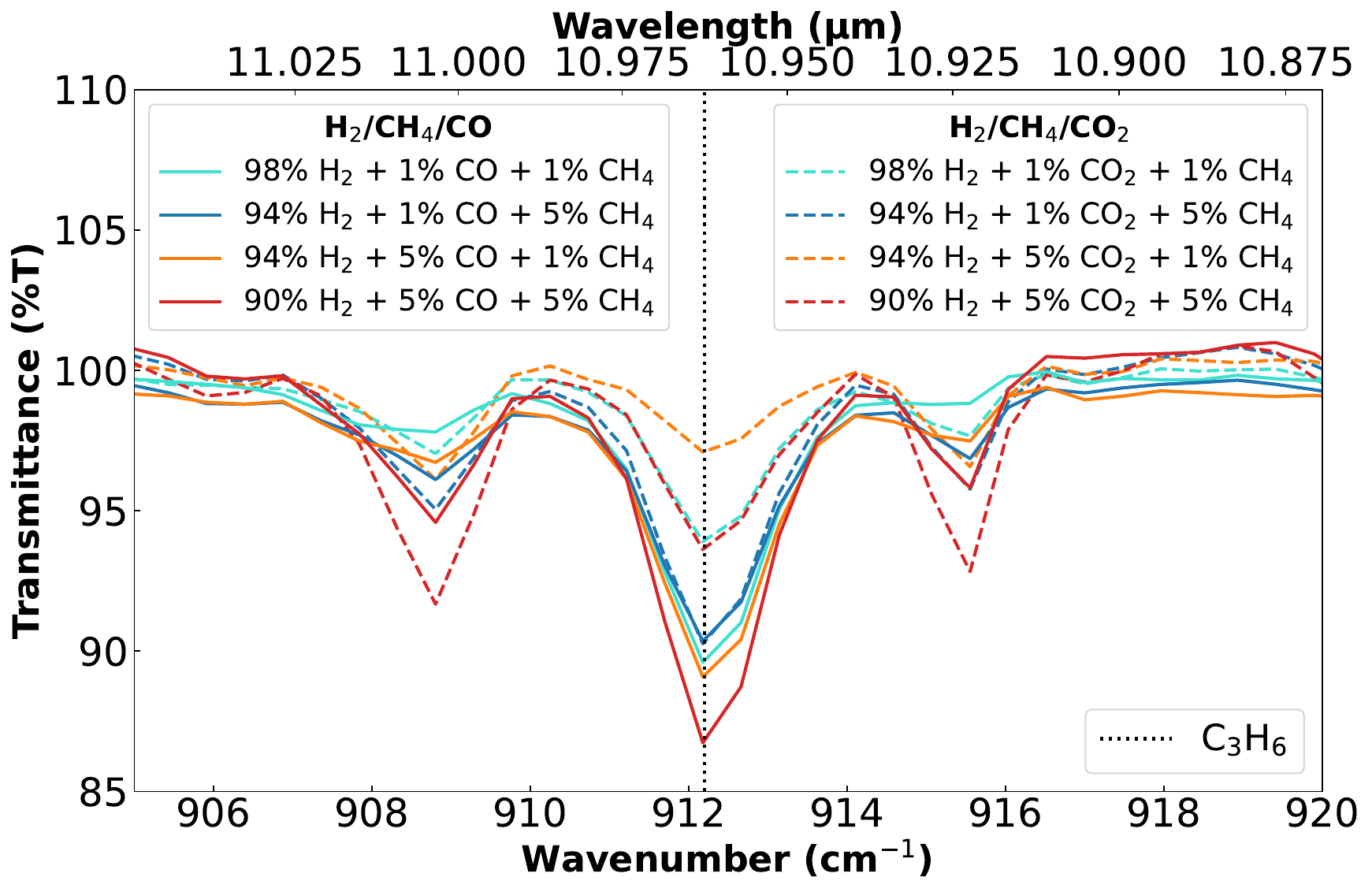}
  \caption{Infrared feature at 912 \cme (10.96 µm), attributed to \ccchhhhhh.}
  \label{IR_912}
  \end{subfigure}
  \caption{Infrared features of the reduced organics.}
  \label{IR_spectra_reduced}
\end{figure*}

\subsection{Abundance and formation trends} 
\subsubsection{Reduced organic compounds}
\label{organic_growth}

\begin{figure*}[ht!]
  \begin{subfigure}{0.49\textwidth}
  \includegraphics[width=\textwidth]{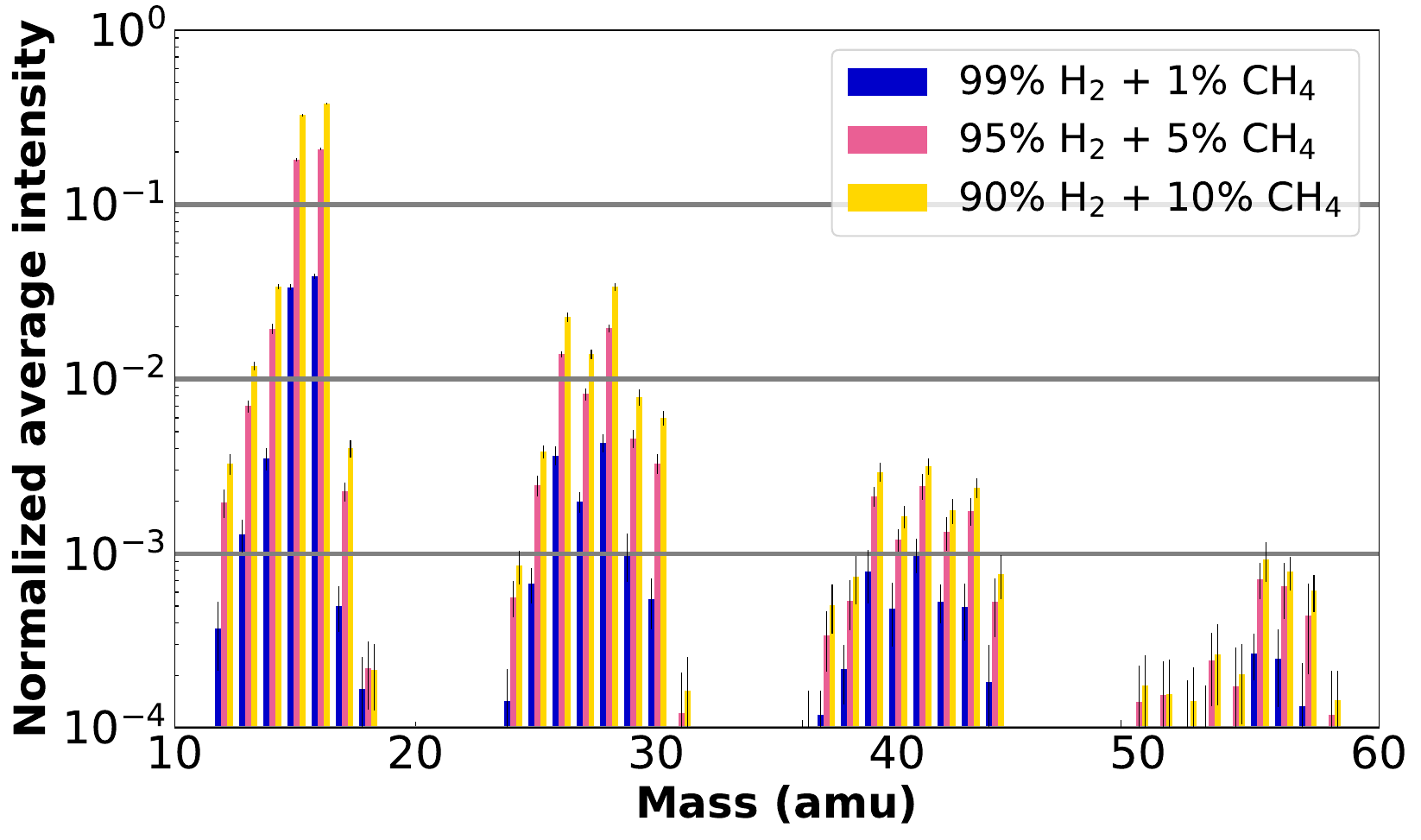}
  \caption{\hh + \chhhh}
  \label{MS_CH4}
  \end{subfigure}

  \begin{subfigure}{0.49\textwidth}
  \includegraphics[width=\textwidth]{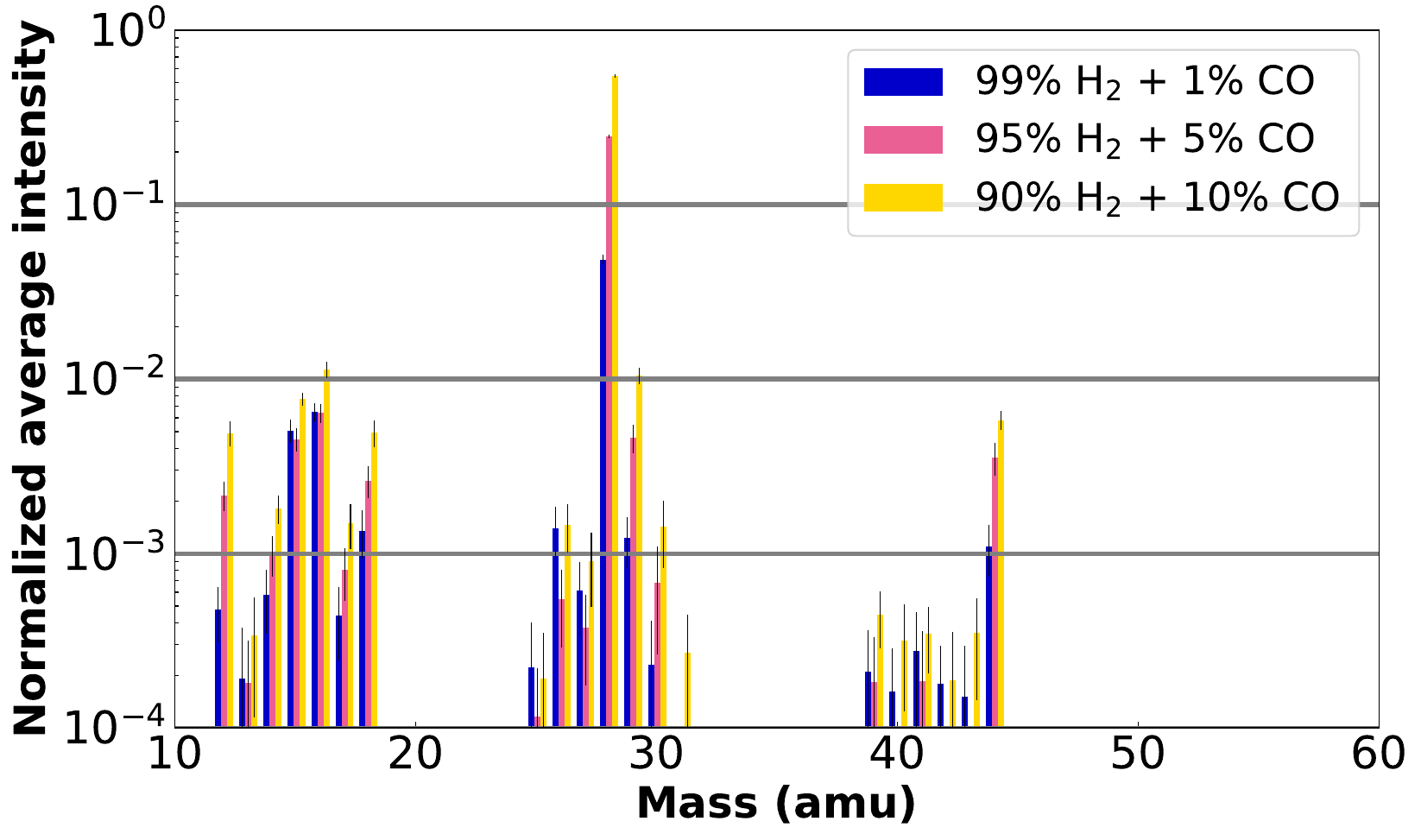}
  \caption{\hh + CO}
  \label{MS_CO}
  \end{subfigure}
  \begin{subfigure}{0.49\textwidth}
  \includegraphics[width=\textwidth]{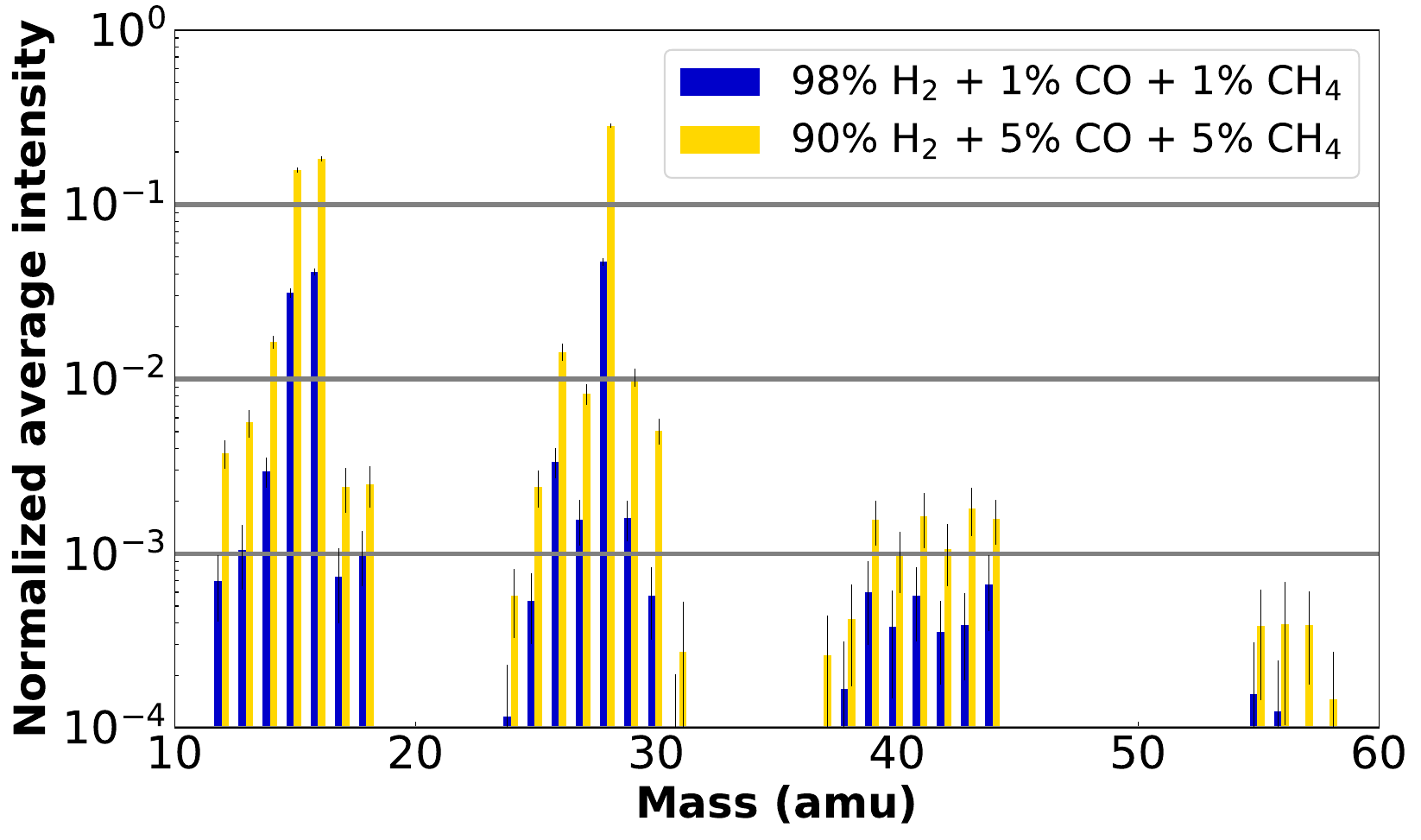}
  \caption{\hh + \chhhh + CO}
  \label{MS_CO_CH4}
  \end{subfigure}
  \begin{subfigure}{0.49\textwidth}
  \includegraphics[width=\textwidth]{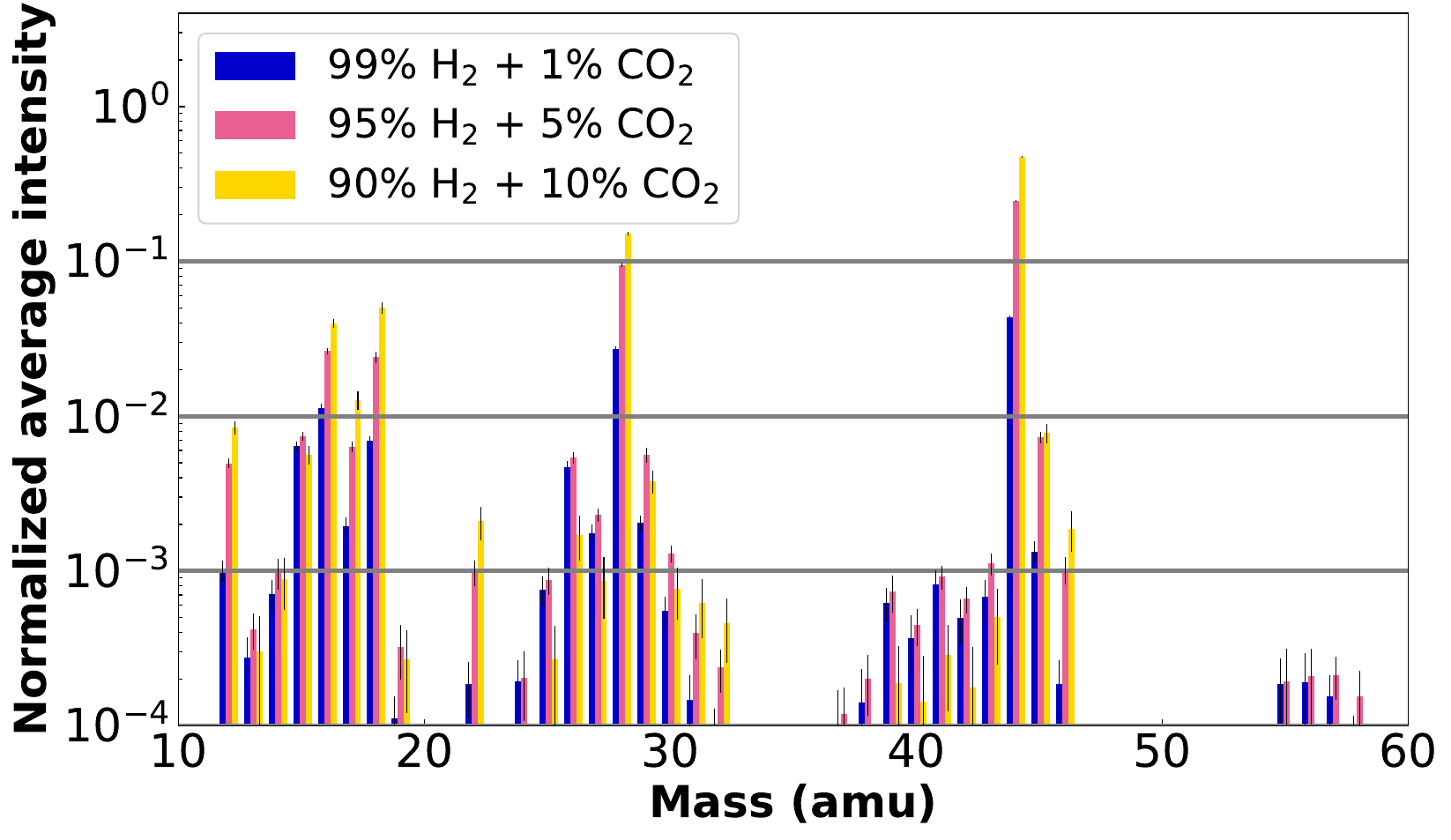}
  \caption{\hh + \coo}
  \label{MS_CO2}
  \end{subfigure}
  \begin{subfigure}{0.49\textwidth}
  \includegraphics[width=\textwidth]{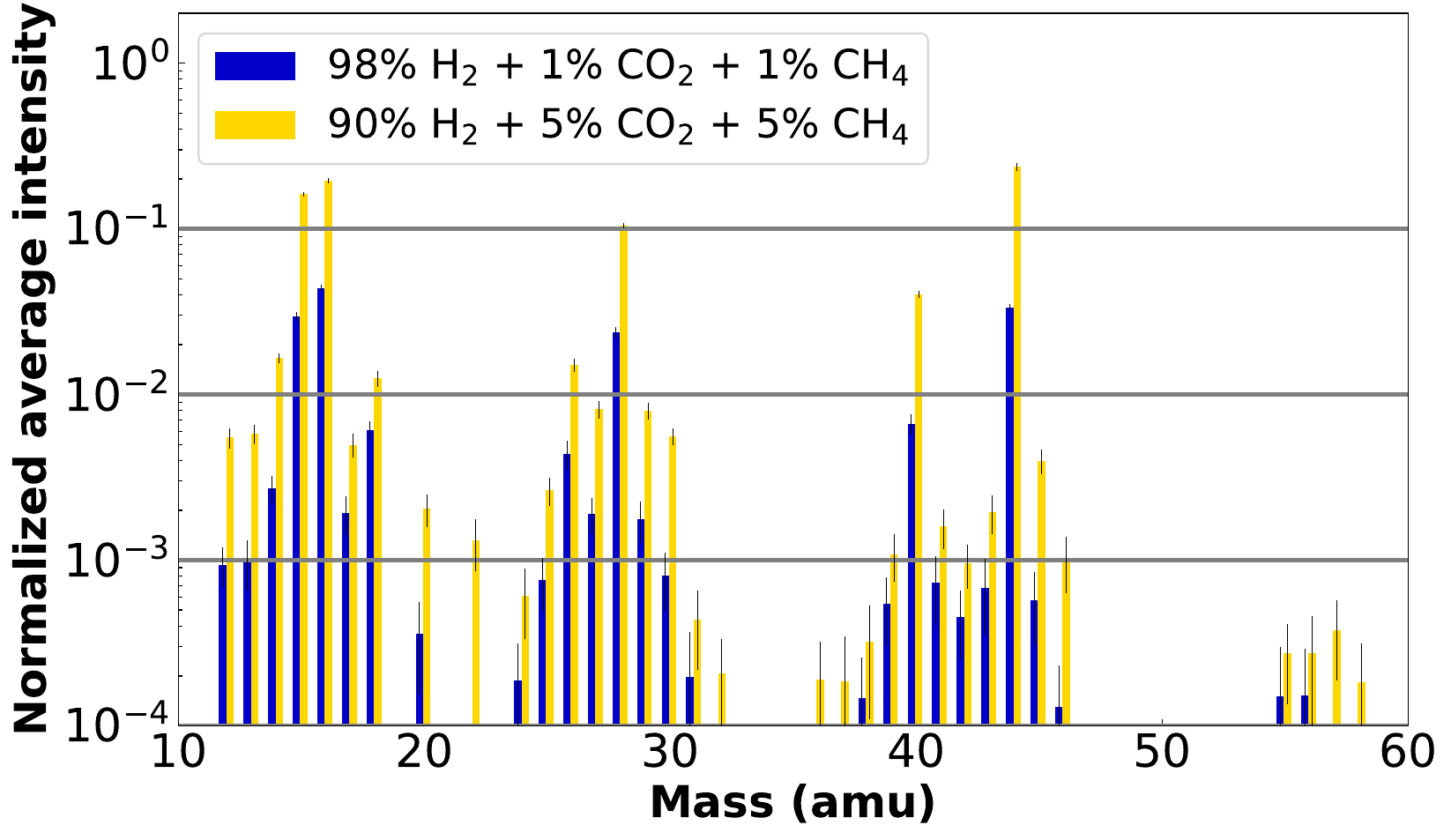}
  \caption{\hh + \chhhh + \coo}
  \label{MS_CO2_CH4}
  \end{subfigure}
  \caption{Mass spectra acquired for different gas mixtures while plasma is ON.}
  \label{MS_spectra}
\end{figure*}

\begin{figure*}[ht!]
  \begin{subfigure}{0.49\textwidth}
  \includegraphics[width=\textwidth]{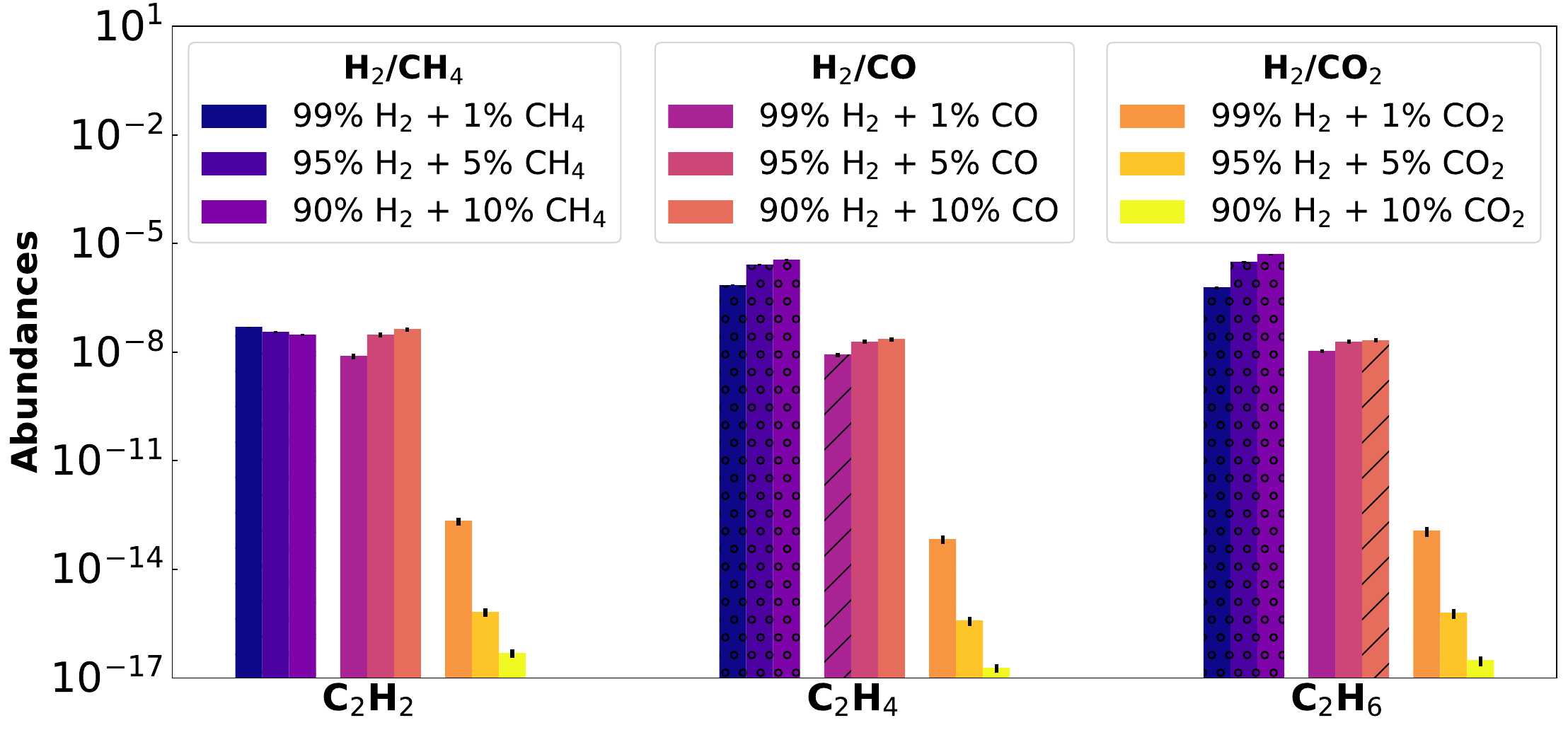}
  \caption{Reduced versus oxidized mixtures}
  \label{reduced_oxizedatm}
  \end{subfigure}
  \begin{subfigure}{0.49\textwidth}
  \includegraphics[width=\textwidth]{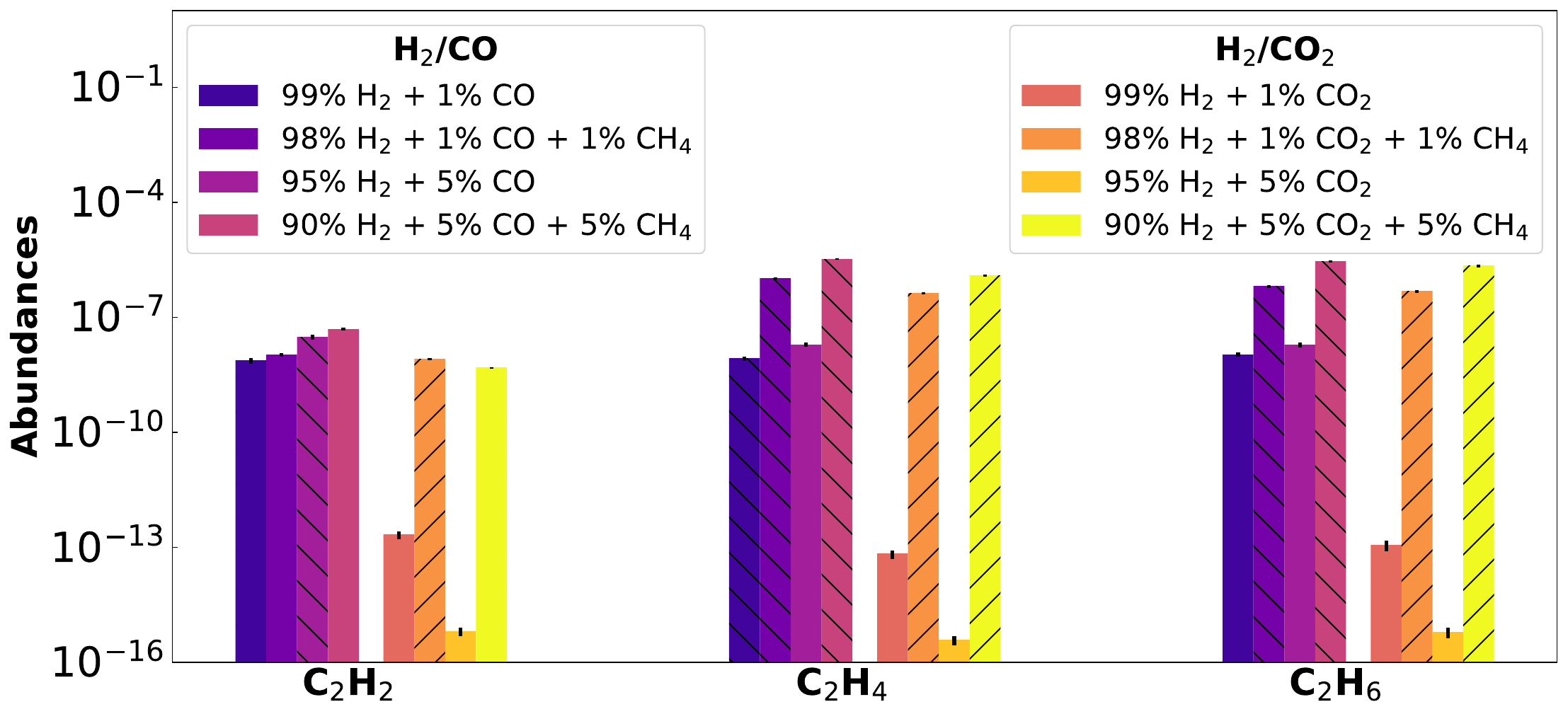}
  \caption{CO-rich versus \coo-rich mixtures}
  \label{reduced_oxvscompatm}
  \end{subfigure}
  \begin{subfigure}{0.49\textwidth}
  \includegraphics[width=\textwidth]{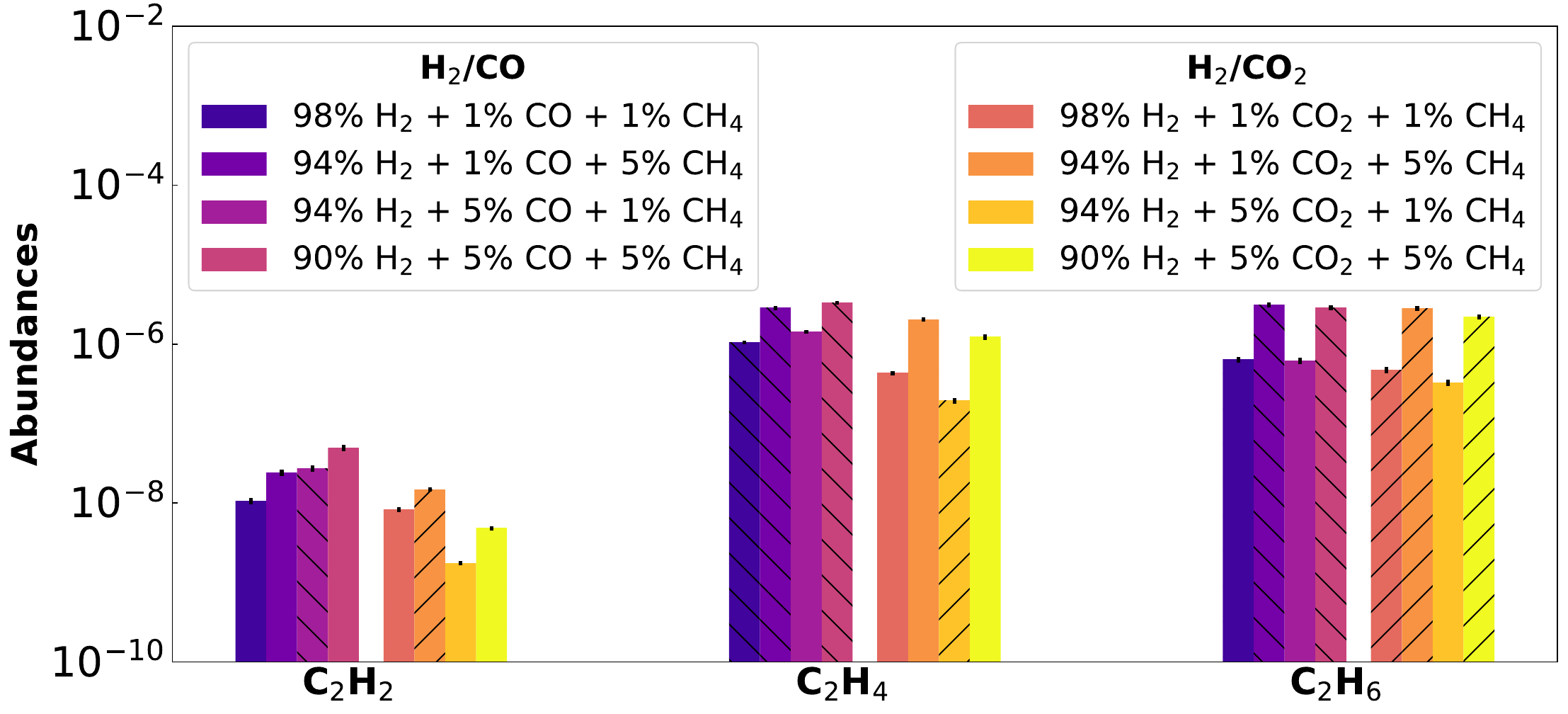}
  \caption{\chhhh/CO versus \chhhh/\coo mixtures}
  \label{reduced_compatm}
  \end{subfigure}
  \begin{subfigure}{0.49\textwidth}
  \includegraphics[width=\textwidth]{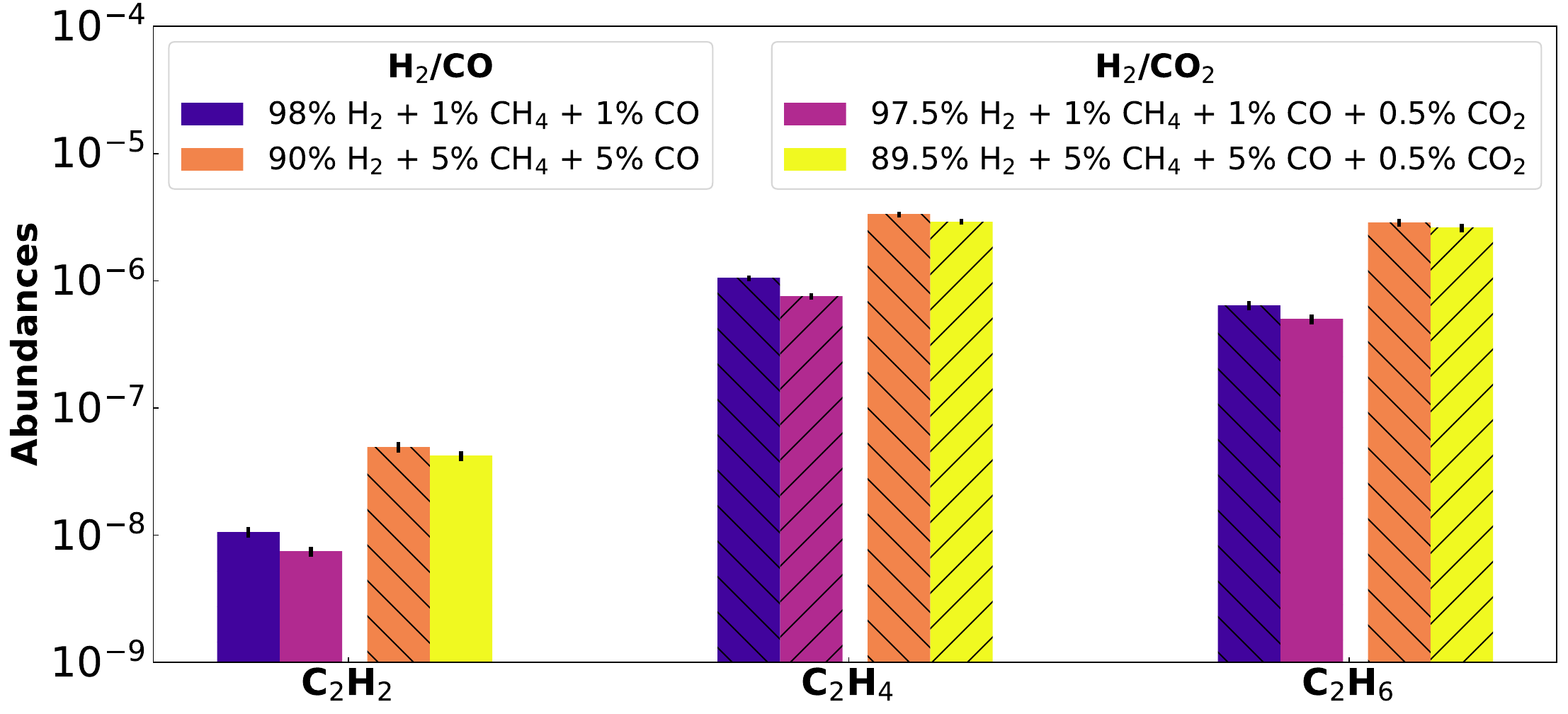}
  \caption{\chhhh/CO versus \chhhh/CO/\coo mixtures}
  \label{reduced_comp3atm}
  \end{subfigure}
  \caption{Predicted abundance values with 0D simulations of the two-carbon atoms reduced organics in the different gas mixtures.}
  \label{reactorui_reduced}
\end{figure*}

The formation of reduced organics is observed under all experimental conditions. We noted the appearance of carbon chain features both in MS and IR as presented in section \ref{sec:identification}, shown in Figures \ref{carbon_chains_ms} and \ref{IR_spectra_reduced}, and detailed in Table \ref{tab: features}. 

The organic growth is most significant in reduced mixtures, and it increases with methane concentration and hence with metallicity, as expected and already highlighted by previous experimental studies \citep{Horst_2018}. The mass spectra show an increase in the intensity of all peaks with increasing methane content, as shown in Figure \ref{MS_CH4}. We observed peak groups characteristic of compounds comprising up to four carbon atoms (C4), at m/z 50-60. Although it is difficult to quantify the species precisely, the relative intensities of the peaks relating to C2 compounds appear to be of comparable magnitudes, reflecting similar concentrations. Based on the 0D simulations performed with ReactorUI, we observe that the abundance of \cchh is approximately two orders of magnitude lower than that of \cchhhh and \cchhhhhh, and decreases with increasing \chhhh concentration (Figure \ref{reduced_oxizedatm}). In contrast, the abundance of \cchhhh and \cchhhhhh increase with increasing methane concentration. 

The formation of reduced organics also occurs in oxidized mixtures, although it is more limited than in reduced mixtures, as previously highlighted in particular for \coo-rich mixtures \citep{Trainer_2004}. It is observed in MS, as we find lower intensities of C2 (m/z 20-30) and C3 (m/z 30-40) specific peak groups in CO and \coo-rich mixtures (Figures \ref{MS_CO} and \ref{MS_CO2}) than in reduced mixtures (Figure \ref{MS_CH4}). In Figures \ref{MS_CO} and \ref{MS_CO2}, we do not observe the C4 specific peak groups at m/z 50-60 that were present in Figure \ref{MS_CH4}, reflecting inhibited organic growth. No selectivity is observed in oxidized mixtures between the C2 compounds, with simulated abundance values similar for \cchh, \cchhhh and \cchhhhhh (Figure \ref{reduced_oxizedatm},\ref{reduced_oxvscompatm}). 

The organic growth is predicted by the 0D-simulations to be more efficient with CO-rich mixtures than with \coo-rich mixtures, as shown in Figure \ref{reduced_oxizedatm} where we see that the abundance values of two-carbon chains are higher with CO than with \coo. This is not obvious with the MS results presented in Figures \ref{MS_CO} and \ref{MS_CO2}, where we see similar relative intensities of the characteristic m/z ratios of compounds C2 and C3. However this is supported by the IR data, as we observe the appearance of species absorption peaks of three-carbon chains only in CO-rich mixtures (CO and \chhhh/CO mixtures). For example, we only see the 912 \cme (10.96 µm) peak attributed to \ccchhhhhh clearly appearing in \chhhh/CO mixtures, as shown in Figure \ref{IR_912} (solid lines). 

For CO-rich mixtures, the organic growth increases with CO concentration. This trend is apparent in the mass spectra in Figure \ref{MS_CO}, where we see the intensities of the C2 and C3 peaks increase with CO concentration. This is confirmed by the 0D simulations performed with ReactorUI, as shown in Figure \ref{reduced_oxizedatm}. The predicted abundance of the C2 compounds is higher at a concentration of 10\% of CO than at a concentration of 1\%.

For \coo-rich mixtures, this trend is reversed: the organic growth is limited at higher \coo concentrations. In the mass spectra, the decrease in the m/z 26 peak — characteristic of \cchh, at a \coo concentration of 10\% — reflects this limitation. The 0D simulations support this trend, with the predicted abundance values of C2 compounds being higher at a concentration of 1\% \coo than at a concentration of 10\%, as shown in Figure \ref{reduced_oxizedatm}. 

When \hh, \chhhh and an oxidized compound — CO or \coo — are merged, the efficiency of organic growth is a hybrid of the two previous cases. It is increased compared to pure oxidized mixtures, and slightly decreased compared to pure reduced mixtures. In the mass spectra, characteristic peak groups of compounds C2 and C3 are seen in Figures \ref{MS_CO_CH4} and \ref{MS_CO2_CH4}, at higher intensities than in the oxidized mixtures presented above, but lower than in the reduced mixtures. Figure \ref{reduced_oxvscompatm} shows the effect of adding methane to the atmospheres, particularly for the formation of \cchhhh and \cchhhhhh: the gain is two to six orders of magnitude, selectively favoring the formation of \cchhhh and \cchhhhhh over \cchh. The more oxidized the composite atmosphere is, the less efficient the formation of reduced organics is. This relationship is visible in Figure \ref{reduced_compatm}, where the predicted abundance values for the 94\% \hh + 5\% [ CO or \coo] + 1\% \chhhh mixtures are the lowest. We also observe in Figure \ref{reduced_comp3atm} that when 0.5\% \coo is added to \chhhh/CO mixtures, the reduced organics abundance values decrease.

\subsubsection{Oxidized organic compounds}
\begin{figure*}[ht!]
  \begin{subfigure}{0.49\textwidth}
  \includegraphics[width=\textwidth]{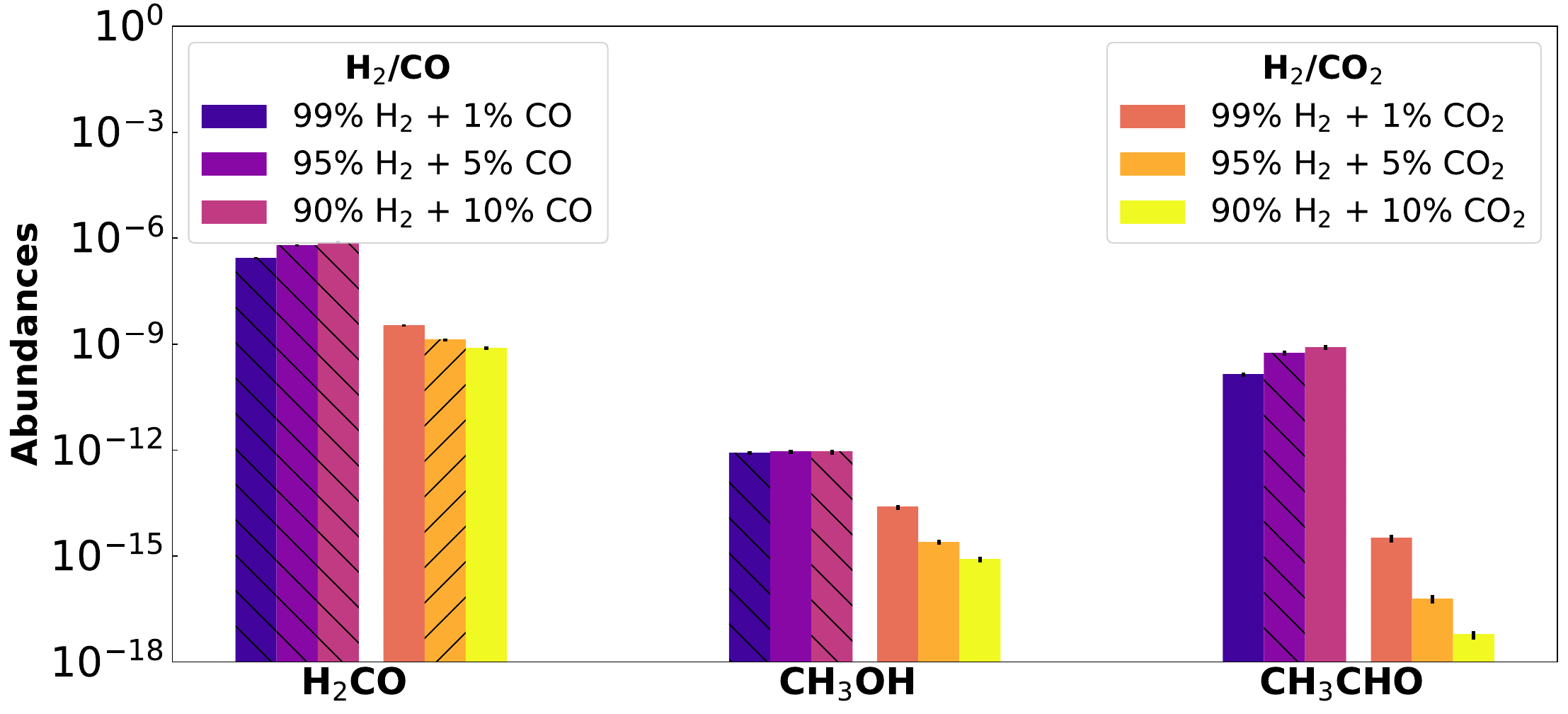}
  \caption{Oxidized mixtures}
  \label{oxidized_oxizedatm}
  \end{subfigure}
  \begin{subfigure}{0.49\textwidth}
  \includegraphics[width=\textwidth]{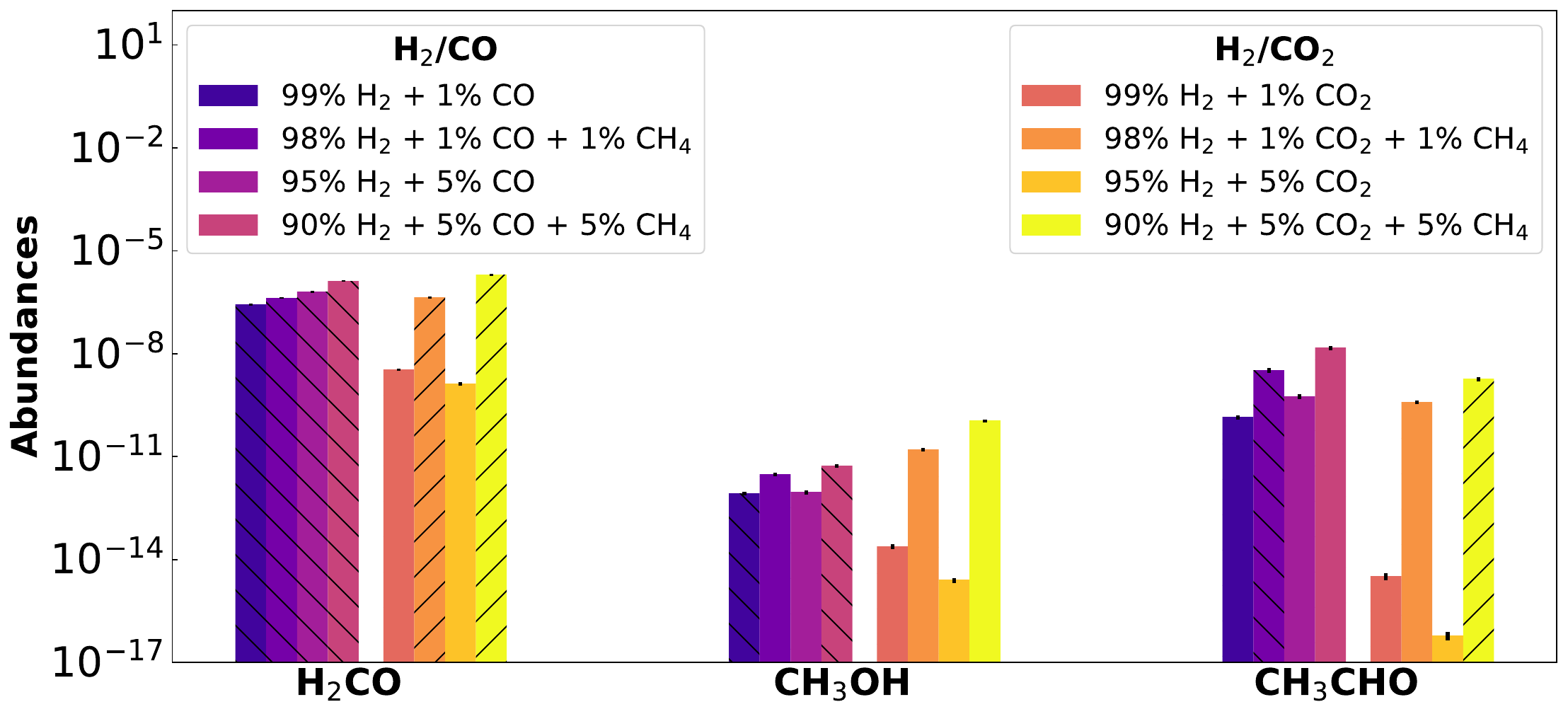}
  \caption{CO-rich versus \coo-rich mixtures}
  \label{oxidized_oxvscompatm}
  \end{subfigure}
  \begin{subfigure}{0.49\textwidth}
  \includegraphics[width=\textwidth]{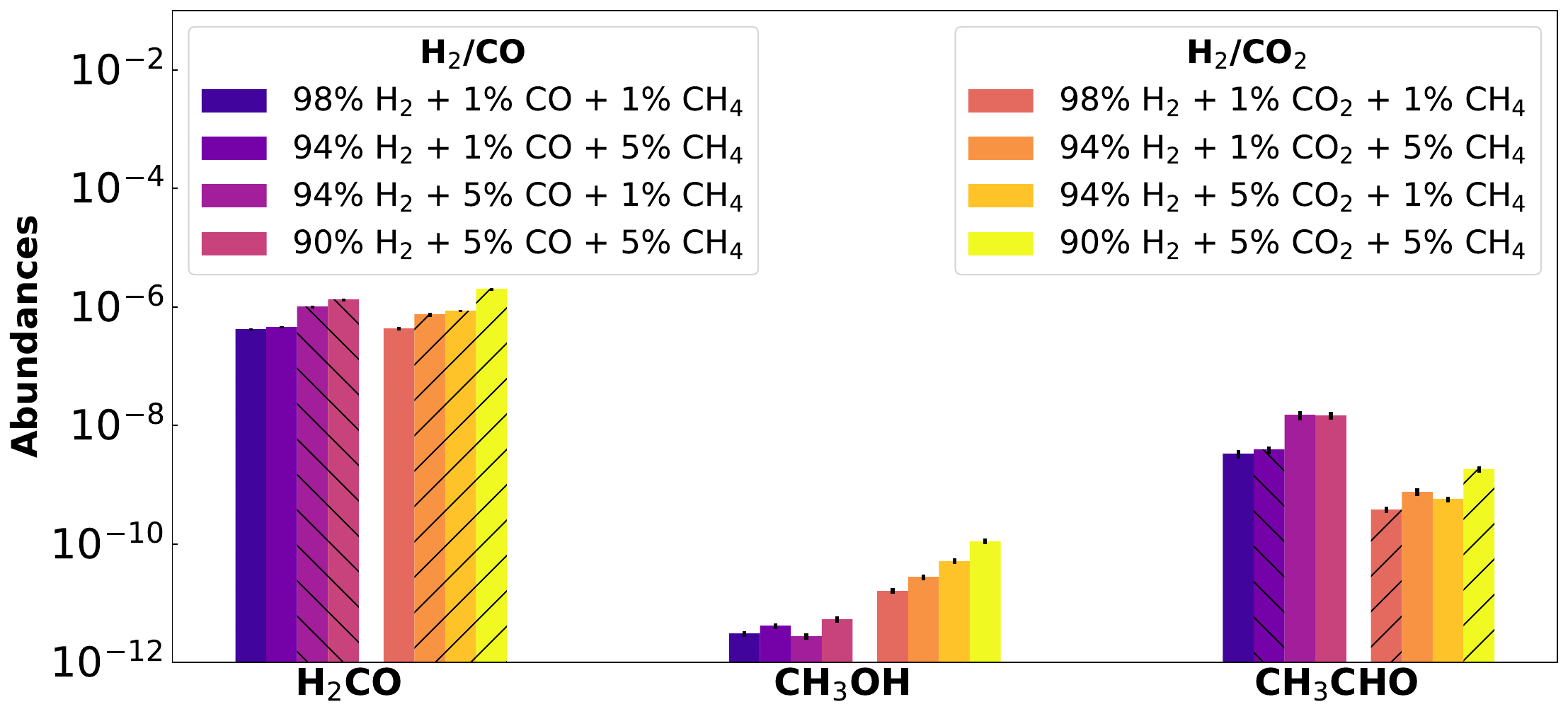}
  \caption{\chhhh/CO versus \chhhh/\coo mixtures}
  \label{oxidized_compatm}
  \end{subfigure}
  \begin{subfigure}{0.49\textwidth}
  \includegraphics[width=\textwidth]{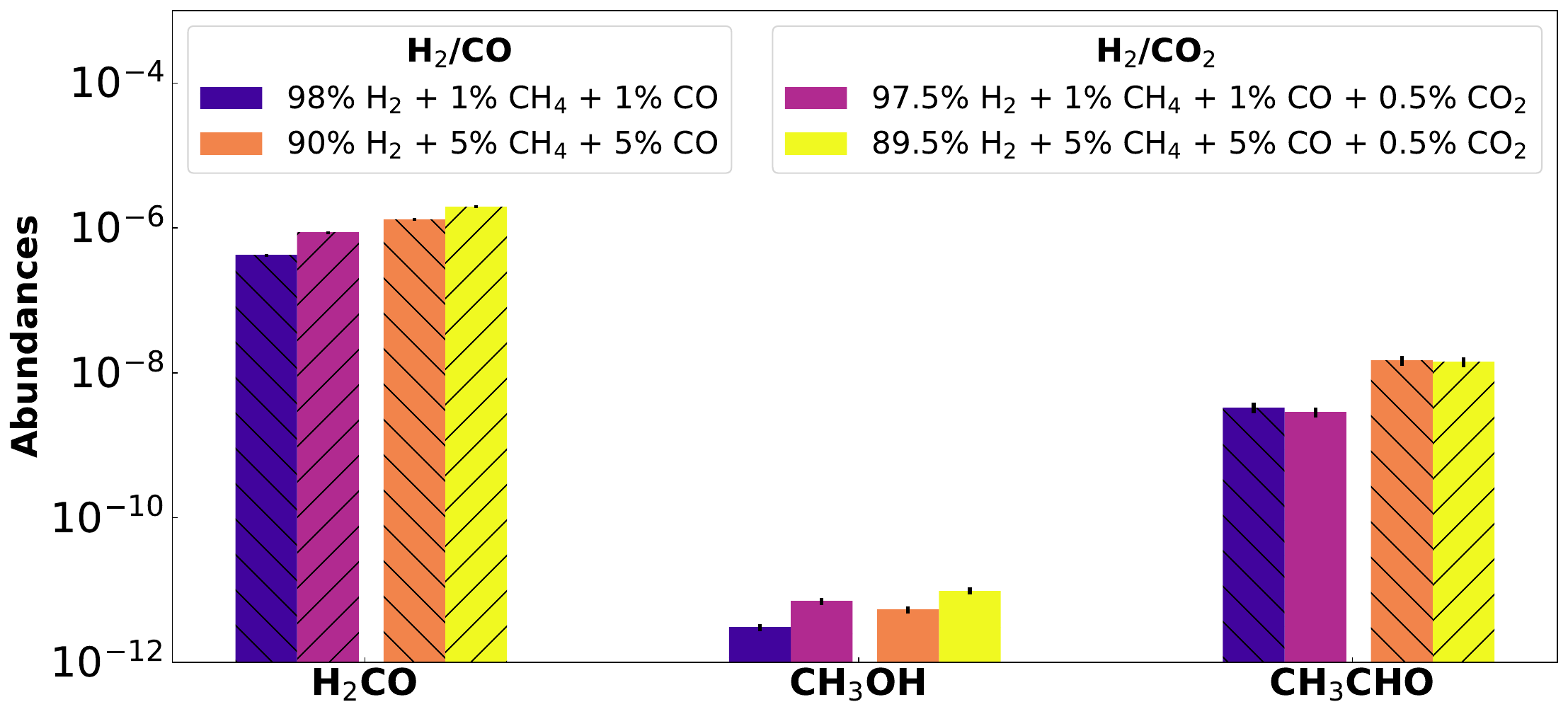}
  \caption{\chhhh/CO versus \chhhh/CO/\coo mixtures}
  \label{oxidized_comp3atm}
  \end{subfigure}
  \caption{Predicted abundance values with 0D simulations of the oxidized organics in the different gas mixtures.}
  \label{reactorui_oxidized}
\end{figure*}

The formation of the oxidized organic compounds methanol, formaldehyde and acetaldehyde is identified through IR features shown in Figure \ref{IR_spectra_oxidized}, only in the atmospheres with CO or \coo. The trapping method, which concentrates the products so that they can be seen in IR, induces a bias that prevents robust quantification directly from the IR results. However, the 0D model allows us to extract trends for the abundance of these oxidized organic compounds, as a function of gas mixtures, shown in Figure \ref{reactorui_oxidized}. 

Figure \ref{oxidized_oxizedatm} shows that the relative abundance values of the three oxidized compounds for CO-rich mixtures are always higher than for \coo-rich mixtures. It also highlights that the growth of these compounds is limited with higher concentrations of \coo, while it is enhanced with high concentrations of CO. The most abundant oxidized organic product observed with 0D simulations is \hhco (Figure \ref{reactorui_oxidized}). It is produced in the same abundance range as reduced organic compounds such as \cchhhh and \cchhhhhh, and even above the abundance of \cchh (Figure \ref{reduced_compatm}).  \chhhoh is much less produced — below the ppb range. 

When \chhhh is added to the atmosphere, the production of oxidized organics is significantly increased (Figure \ref{oxidized_oxvscompatm}), and enhanced at higher concentrations of both CO and \coo. In this case, we observe that the formation of \hhco and \chhhoh becomes more efficient with \coo than with CO. We observe in Figure \ref{IR_spectra_oxidized} that for a given concentration (same color), the intensity of the feature always appears to be greater in \hh/\chhhh/\coo mixtures (dotted lines) than in \hh/\chhhh/CO mixtures (solid lines). This is confirmed by the results of the 0D simulations, where higher abundance values for \hhco and \chhhoh are observed in \coo-rich mixtures than in those that are CO-rich (Figure \ref{oxidized_compatm}). On the other hand, the abundance values of \chhhcho remains higher with CO than with \coo. The combination of \chhhh, CO and \coo increases the abundance of all oxidized organics of about one order of magnitude (Figure \ref{oxidized_comp3atm}).

\subsection{Formation mechanisms}
From the 0D simulations performed with ReactorUI, we can determine the key reactions involved in the production and destruction of each chemical compound. This allows us to reconstruct the chemical networks involving the molecules that interest us in our study.
\subsubsection{Reduced organic compounds}
\begin{figure*}[ht!]
  \begin{subfigure}{0.43\textwidth}
  \includegraphics[width=\textwidth]{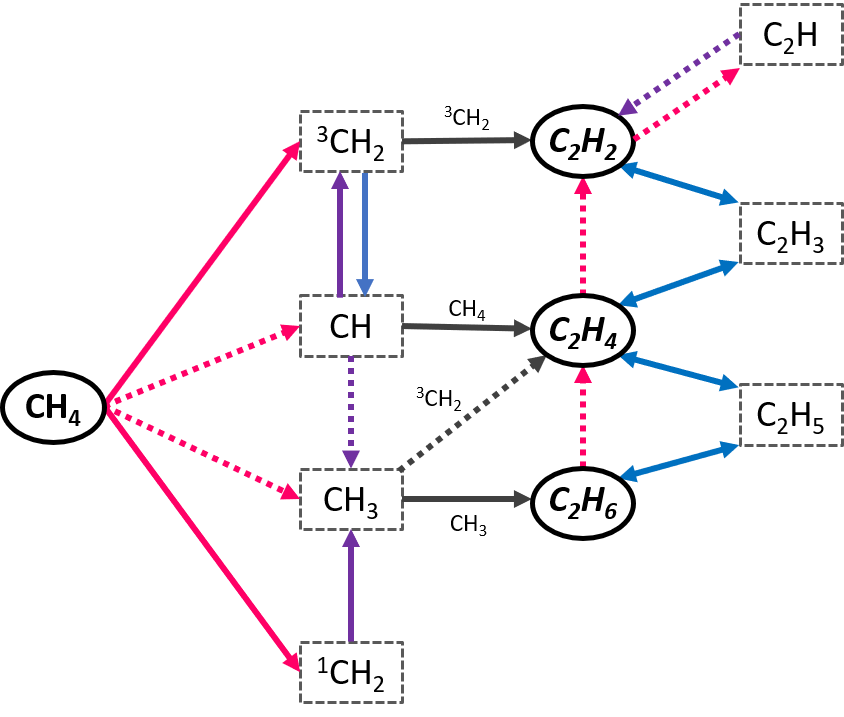}
  \caption{}
  \label{path_reduced}
  \end{subfigure}
  \begin{subfigure}{0.59\textwidth}
  \includegraphics[width=\textwidth]{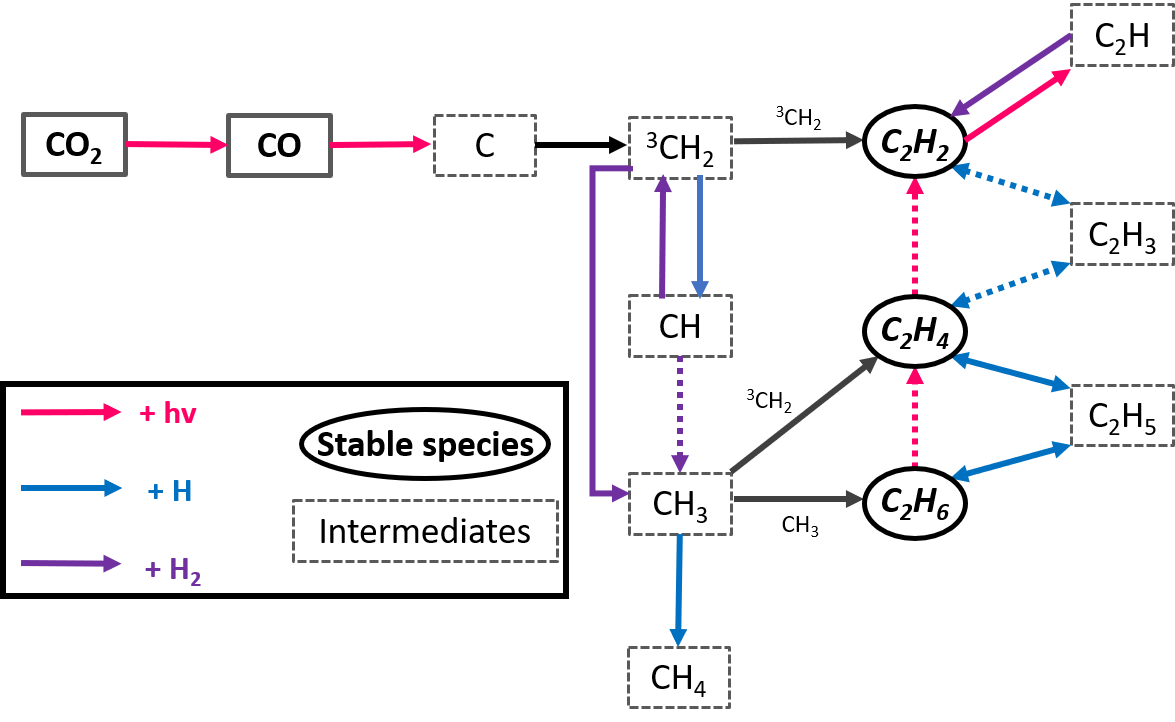}
  \caption{}
  \label{path_oxidized}
  \end{subfigure}
  \caption{Formation pathways of reduced organics with main (solid arrows) and minor (dotted arrows) reactions started from a reduced reactant (\chhhh \ref{path_reduced}) or an oxidized one (CO or \coo \ref{path_oxidized}).}
  \label{pathways_reduced}
\end{figure*}

Figure \ref{path_reduced} shows the formation pathway of the C2 compounds from the reactant \chhhh in reduced mixtures. From the dissociation of \chhhh, several intermediates are produced, which recombined to form the C2 compounds. \cchhhhhh is mainly formed by the combination reaction of 2 \chhh compounds, \cchhhh is mainly produced by the reaction of CH with \chhhh, and \cchh mainly comes from the combination of two triplet methylene radicals $^3$CH$_2$. The radical reaction required for the production of \cchh — which is approximately three orders of magnitude slower than the combination reactions forming \cchhhh and \cchhhhhh — could explain its lower abundance predicted by 0D simulations. The relative contributions of these reactions in the formation of the reduced organics remains more or less the same as the concentration of \chhhh increases. That is coherent with the fact that the higher the proportion of \chhhh in the atmosphere — and therefore the higher the metallicity — the higher the final concentrations of reduced organic compounds. 
 
This chemical reactions network (Figure \ref{path_reduced}) is consistent with the well-known methane chemistry and hydrocarbon formation studied in the reduced atmospheres of our Solar System, such as Titan \citep{Nixon_2024}, Jupiter \citep{Moses_2005} or Saturn \citep{Moses_2000}, as well as with predictions made for exoplanets with \hh dominant atmospheres \citep{Adams_2022}. This validates our approach, and gives us confidence for the study of our more oxidized atmospheric ranges.

\begin{figure*}[ht!]
  \centering
  \includegraphics[width=0.8\textwidth]{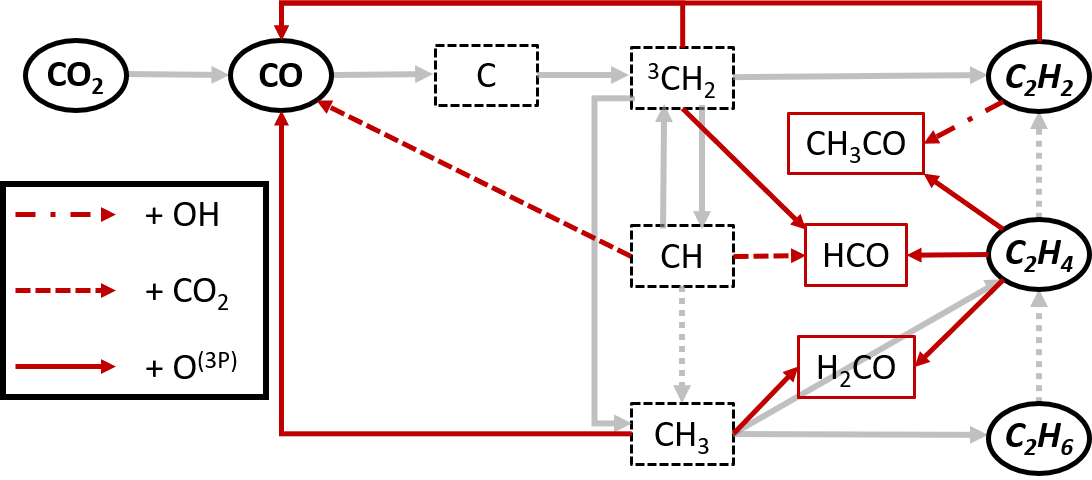}
  \caption{Main pathways of loss through oxidation of reduced organic compounds (red) superimposed on their formation pathways (light gray) presented in Figure \ref{path_oxidized}.}
  \label{loss_pathways_reduced}
\end{figure*}

Figure \ref{path_oxidized} shows the formation pathway of reduced organics from CO and \coo. Organic growth is inhibited compared to reduced mixtures, as has been previously studied in the literature — particularly in the \coo case by \cite{Trainer_2004}. It requires passing through more intermediates than in the reduced atmospheres, which limits the formation of reduced organic compounds. The formation of \cchhhh and \cchhhhhh requires the prior formation of the \chhh intermediate, which in turn is produced by two consecutive intermediates (C and $^3$CH$_2$) after CO dissociation. This long reaction path explains the lower abundance of hydrocarbons compared to the reduced atmosphere. Furthermore, the dissociation of CO, which provides a source of carbon atoms for organic growth, is at the same time a source of exited oxygen radicals, mainly in the \op state, which will oxidize some of the carbonate radical intermediates and reduced organics (Figure \ref{loss_pathways_reduced}). For example, $^3$CH$_2$ reacts with \op to reform CO. However, as the concentration of CO in gas mixtures increases, there is an overall rise in the formation of reduced organic compounds, showing that with CO, the effect of increasing the carbon source dominates over the loss of intermediates by oxidation for the formation of the organics. 

For \coo-rich mixtures, the difference in efficiency in the formation of reduced organic compounds compared to CO-rich mixtures can be linked with the fact that it first requires the dissociation of \coo before producing the intermediates. This produces CO and an exited oxygen radical, mainly in the \od state. The conversion of \coo into CO is not total, which explains why the source of carbon is less abundant than in the case where CO is a reactant. In addition, the production of oxygen radicals is doubled, as a first radical is produced from the dissociation of \coo and a second from the dissociation of CO. The oxidation of carbonate intermediates and reduced organics is more important. For example, the reaction of \cchh with OH radical, which is a minor loss reaction in the case of CO, becomes dominant in the case of \coo. This is related to the fact that the formation of OH radicals is more efficient from \coo than from CO, thanks to the following combination reaction, which is an order of magnitude faster with \coo than with CO:
\begin{equation}
  O^1D + H_2 \longrightarrow OH + H .
\end{equation}
The main pathways of loss through oxidation are shown in Figure \ref{loss_pathways_reduced}. These combined effects explain why the abundance of reduced organic substances in \coo-rich gas mixtures is much lower than in CO-rich mixtures, and why the abundance of reduced organic compounds decreases with increasing \coo. 

As \chhhh is a reactant in the \chhhh/CO/\coo mixtures, the reaction pathway presented in Figure \ref{path_reduced} also explains the organic growth. When atmospheres are more oxidized, we can observe the same mechanisms that are distinctive of fully oxidized atmospheres, as developed in Figure \ref{path_oxidized}, but with a limitation in organic growth due to the oxidation of reaction products and intermediates.

In summary, organic growth efficiency is best in \chhhh-rich mixtures, as expected, and is significant in oxidized mixtures, although more limited. In mixtures rich in \coo, oxidative loss reactions become predominant and limit hydrocarbon formation, even at high metallicity values. In CO-rich mixtures, the effect of the carbon source partially compensates for the inhibitory effect of the oxygen source, and we observe that organic growth efficiency increases with metallicity, as in reduced gas mixtures.

\subsubsection{Oxidized organic compounds}
\begin{figure*}[ht!]
  \begin{subfigure}{\textwidth}
  \centering
  \includegraphics[width=\textwidth]{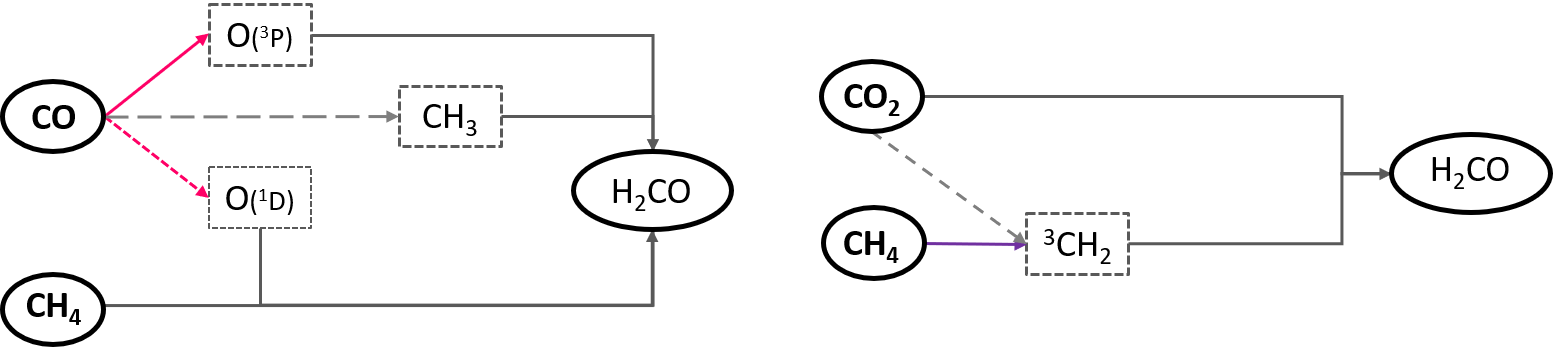}
  \caption{Formation pathways of \hhco}
  \label{pathways_h2co}
  \end{subfigure}
  
  \begin{subfigure}{\textwidth}
  \centering
  \includegraphics[width=\textwidth]{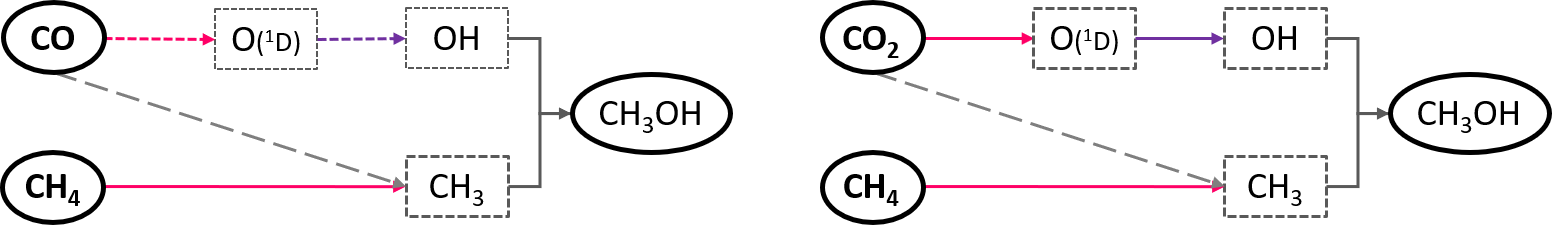}
  \caption{Formation pathways of \chhhoh}
  \label{pathways_ch3oh}
  \end{subfigure}
  
  \begin{subfigure}{\textwidth}
  \centering
  \includegraphics[width=\textwidth]{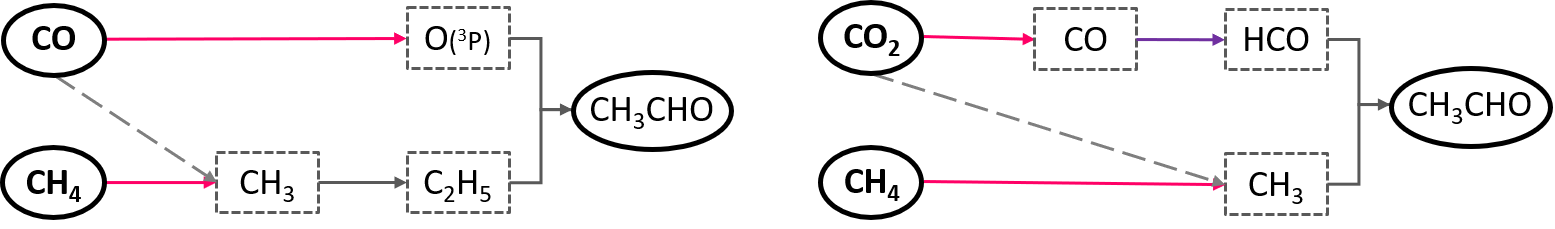}
  \caption{Formation pathways of \chhhcho}
  \label{pathways_ch3cho}
  \end{subfigure}
  \caption{Formation pathways of oxidized organics. The formation pathways of the intermediates \chhh and $^3$CH$_2$ from CO or \coo are detailed in Figure \ref{pathways_reduced} and are simplified here by long dotted gray arrows. The main (solid arrows) and minor (short dotted arrows) reactions are represented.}
  \label{pathways_oxidized}
\end{figure*}

The main formation pathways of \hhco and \chhhcho are different depending on the oxidized reactant, as shown in Figure \ref{pathways_h2co} and \ref{pathways_ch3cho}, while for \chhhoh it is similar for CO and \coo. All these reaction paths involve reduced organic intermediates, such as \chhh and $^3$CH$_2$, for which the formation efficiency in CO or \coo mixtures is described in section \ref{organic_growth}. This explains the formation trends observed, i.e., formation efficiency is better with CO than with \coo, enhanced at high CO concentrations and limited at high \coo concentrations. The formation of \chhhoh and \chhhcho in mixtures of CO and \coo requires at least one more intermediate than the formation of \hhco. This makes the formation of \hhco up to six orders of magnitude faster, explaining its predominance.

The main loss pathways of oxidized organic compounds could provide an additional explanation for the predominance of \hhco over \chhhoh and \chhhcho. Indeed, in all gas mixtures, the loss of \hhco is mainly due to the following reactions: 
\begin{equation}
  H_2CO + H \longrightarrow HCO + H 
\end{equation}
\begin{equation}
  H_2CO + h\nu \longrightarrow CO + H_2 .
\end{equation}
The reaction of \hhco with H is almost two orders of magnitude faster than its photodissociation in the 0D model. HCO and CO are themselves involved as reactants in the formation of \hhco. In this way, \hhco is self-sustaining, which explains its high abundance. In contrast, \chhhoh can be photo-dissociated through 
\begin{equation}
  CH_3OH + h\nu \longrightarrow H_2CO + H_2,
\end{equation}
which forms \hhco again. 

The addition of \chhhh in the gaseous mixture enhances all the productions, since the formation pathways of the reduced intermediates are much shorter starting from the reactant \chhhh than from CO or \coo. With this limiting effect removed, the production of \hhco and \chhhoh from \coo is more efficient than with CO. This is because \coo is a direct reactant in the formation of \hhco, whereas with CO, the intermediate \od is required (Figure \ref{pathways_h2co}), which is a minor product during the dissociation of CO. For the formation of \chhhoh, the same reaction pathway is followed from CO or from \coo (Figure \ref{pathways_ch3oh}), requiring the intermediate \od, a major product during the dissociation of \coo. However, since the formation of \chhhcho from \coo requires one more intermediate than from CO (Figure \ref{pathways_ch3cho}), it remains more efficient with CO than with \coo.

In summary, the efficiency of the formation of oxidized organic compounds is greater in \hh/CO mixtures and increases with metallicity. In \hh/\coo mixtures, formation of oxidized organic compounds is limited and decreases with the metallicity. However, the combination of \chhhh with CO or \coo significantly increases the formation of oxidized products, in correlation with metallicity. The efficiency of \hhco and \chhhoh formation is therefore higher in \chhhh/\coo mixtures than in \chhhh/CO mixtures.

%-----------------------------------------------------------------

\section{Discussion}

\subsection{Roles of methane, carbon monoxide, and carbon dioxide in organic growth}
Methane plays a predominant role in organic growth, as demonstrated in our experiments, in which we observed a positive correlation between gas-phase hydrocarbon production and methane concentration. This is consistent with previous experiments in the literature, which have highlighted the major role of methane in the growth of solid particles (tholins) when \nn/\chhhh Titan-type mixtures are irradiated \citep{Horst_2013,Horst_2018,Moran_2022,Jovanovic_2020, Jovanovic_2021}. The production of the solid phase in Titan-type mixtures is optimized between 3 and 5\% of \chhhh \citep{Sciamma_2010}. In this study, in \hh-dominated mixtures up to 10\% of \chhhh, we observe continuous growth of hydrocarbons in the gas phase, without an obvious optimal concentration. This optimum would possibly occur at higher proportions of \chhhh.

While methane is clearly identified as a major driver of organic growth, the effect of oxidized reactants is not as clearly correlated. At first glance, we identified an inhibitory effect due to the atomic oxygen source represented by the oxidized reactants,  thus limiting the growth of long carbon chains. This is even more pronounced with \coo, which provides twice as much oxygen. This effect has also been identified in the production of solid particles in \nn/\chhhh mixtures, where the addition of CO alters the kinetics of methane, delaying the appearance of solid particles and reducing their production rate \citep{Fleury_2014,He_2017,Moran_2022}.

However, CO and \coo are also sources of carbon and can contribute to organic growth. \cite{Trainer_2006} showed that adding \coo to a \nn/\chhhh mixture could increase particle production up to a \chhhh/\coo ratio of 1 before decreasing. This effect is particularly evident in our \chhhh-free mixtures, where hydrocarbon production is still observed, although at lower levels than in the case where \chhhh is present. The addition of CO to \nn/\chhhh mixtures, although reducing the production rate of tholins, increases their size \citep{Fleury_2014,HorstTolbert_2014}. In our study, we observe a positive correlation between the formation of organic compounds in the gas phase and the concentration of CO. That highlights that with CO, the effect of increasing the carbon source dominates over the loss by oxidation. In the case of a low methane concentration (1\%), \cite{Jovanovic_2020} observed a higher production rate in the presence of 500 ppm CO than in its absence, highlighting the important role that CO can play as a carbon source when methane is less dominant.

Furthermore, the addition of oxygen atoms increases the diversification and organic complexity of photochemical products. We observe the appearance of oxidized organic compounds of prebiotic interest in our CO and \coo-rich mixtures (\chhhoh, \hhco, \chhhcho). It has also been shown that adding CO to Titan-type mixtures improves the reactivity of the gas phase \citep {Yang_2025, Fleury_2014}, leading to the incorporation of oxygen in both the gas and solid phases \citep{He_2017}. Similarly, the addition of \coo leads to more oxidized solid compounds \citep{Trainer_2004}. Since \coo is a more important source of oxygen than CO, one would expect oxygen incorporation to be better. In our study, this corresponds to the observation that oxidized organic compounds are better produced in the gas phase with \coo than with CO when combined with \chhhh and therefore when organic growth is not too limited. For the solid phase, \cite{Drant_2025} observed in \nn/\chhhh-rich exoplanetary haze analogs that the characteristic absorption bands of oxygenated functions (O–H, C=O) are clearly visible in analogs produced with \coo, but not or only very weakly in those produced with CO alone.

\subsection{Photochemistry of hydrocarbons in temperate methane-rich exoplanet atmosphere}
Experiments conducted in reduced gas mixtures reveal similar formation rates for C2 compounds, preventing any definitive conclusions about selectivity. Simulated abundance using the 0D model also predicts comparable values for these molecules, but with a slight selectivity observed for \cchhhhhh and \cchhhh relative to \cchh.
This contrasts with the hydrocarbon photochemistry observed in the gas giants Jupiter and Saturn, where detected abundance is depleted in \cchhhh and enriched in \cchhhhhh and \cchh \citep{Moses_2000,Moses_2005,Nixon_2010}. These planets have hydrogen-dominated atmospheres with significant helium concentrations and a small fraction of methane (around 0.1\%). They are cooler — around 150 K at 1 mbar. 

The low abundance of \cchhhh is explained by its high photodissociation rate at Lyman-$\alpha$, producing \cchh, and by its reaction with H, which recycles it into \chhhh \citep{Gladstone_1996}. In contrast, \cchhhhhh is shielded by methane and thus remains stable at Lyman-$\alpha$. 
In our study, the energy distribution of electrons in our plasma covers a much wider range than that of photons in planetary atmospheres, which are largely concentrated around Lyman-$\alpha$. Consequently, the wavelength selectivity of photodissociation is not reproduced, which likely contributes to the differences observed in the relative abundance of C2 compounds between our experiments and planetary atmospheres.

Additionally, \cchh, as a highly unsaturated compound, would normally be unstable in an \hh-dominated environment. However, it is efficiently recycled via the recombination reaction
\begin{equation}
  C_2H + H_2 \longrightarrow C_2H_2 + H .
\end{equation}
We noticed that this reaction is negligible in our reduced gas mixtures, whereas it becomes significant in oxidized mixtures. Since CO and \coo are sources of carbon but create a less reducing environment than \chhhh, this increases the degree of unsaturation of the products. This has also been demonstrated for the production of solid products \citep{He_2017,Yang_2025}. This corresponds to the fact that we observe a depletion of \cchh in mixtures containing only \chhhh, whereas there is no apparent selectivity in mixtures containing CO and \coo.

Temperature differences may also play a role in C2 relative abundance. \cite{Adams_2022} suggests that warmer \hh-rich atmospheres, compared to Titan, favor dynamic chemical cycles involving \cchh, \cchhh, and \cchhhh through reactions with atomic hydrogen. These processes can reduce \cchhhhhh and \cchh concentrations while increasing the relative abundance of \cchhhh.

\subsection{Detectability of photoproducts}
\subsubsection{Hydrocarbons}
The C2 compounds appear to be relatively abundant and efficiently produced in our experiments, which are dominated by out-of-equilibrium chemistry. Similarly, a 1D non-equilibrium chemical model fitting the JWST observational data of K2-18 b predicts that \cchh, \cchhhh and \cchhhhhh could reach volume mixing ratios of up to a few percent in the upper atmosphere (at 1 mbar) \citep{Jaziri_2025}. 
\cchh and \cchhhh are in principle detectable with JWST, using combined observations from the NIRSpec/G395 and MIRI LRS instruments, provided their mixing ratios exceed 10$^{-6}$ \citep{Gasman_2022}. This suggests that the detection of hydrocarbons in the atmosphere of K2-18 b is plausible if methane is abundant and predominant over carbon dioxide, as predicted by \cite{Wogan_2024,Schmidt_2025,Jaziri_2025,Hu_2025}, and if photochemistry is sufficiently active. The 3.4 and 6.9 µm features observed in the NIRSpec and MIRI data could plausibly be attributed to \cchhhhhh \citep{Luque_2025}, providing an alternative to their attribution to DMS (dimethyl sulfide) or DMDS (dimethyl disulfide) by \cite{Madhusudhan_2025}. In summary, two-atom hydrocarbons are good candidates for detection with current JWST observations and future high-resolution ground-based observations, both in terms of expected abundance and photochemical plausibility. It is necessary to consider them before considering more complex species with a variety of potential sources.

\subsubsection{Oxidized organics}
Our complementary experimental approach combining MS and IR spectroscopy enables robust identification of \hhco, \chhhoh, and \chhhcho formation in oxidized gas mixtures. \cite{Jaziri_2025} predict that \hhco, \chhhoh, and \chhhcho could reach volume mixing ratios of up to a few parts per million (10$^{-6}$) in the upper atmosphere of K2-18 b. However, detectability studies with JWST indicate that species such as \hhco and \chhhoh require volume mixing ratios exceeding 10$^{-5}$ to produce measurable features in \hh-dominated atmospheres \citep{Huang_2022,Zhan_2022}. Their detection in the atmosphere of K2-18 b therefore seems unlikely given current sensitivity limits. Nevertheless, our results suggest that in other \hh-rich exoplanets with a higher CO and/or \coo content, the formation of oxidized organic species may be particularly efficient. In such environments, these molecules could form more abundantly and potentially reach detectable levels with JWST or with future high-resolution ground-based instruments on the Extremely Large Telescope (ELT), such as ANDES (ArmazoNes high Dispersion Echelle Spectrograph, \citealt{ANDES_2025}) and PCS (Planetary Camera and Spectrograph, \citealt{PCS_2021}).

%--------------------------------------------------------------------

\section{Conclusion}

We studied out-of-equilibrium chemistry in \hh-dominant atmospheres typical of temperate sub-Neptunes by combining experimental and numerical simulations. This approach complements previous laboratory studies by enabling more accurate molecular identification and clearer elucidation of reaction pathways. Our experiments, which are based on the combined use of MS and IR spectroscopy, enabled the reliable detection of reduced (\cchh, \cchhhh, \cchhhhhh) and oxidized (CO, \coo, \hho, \hhco, \chhhoh, \chhhcho) species in gas mixtures covering a wide range of C/O ratios and metallicities. The use of the 0D photochemical model allowed us to extract abundance trends and identify formation pathways.

Hydrocarbon compounds are produced in methane-rich atmospheres in particular. The efficiency of this organic growth increases with metallicity in \chhhh-rich and CO-rich atmospheric analogs. In \coo-rich atmospheric analogs, oxidative loss reactions become predominant and limit the formation of hydrocarbons. The formation of hydrocarbons from methane can be explained by reaction pathways that have already been well studied in the atmospheres of the Solar System, whereas their formation from oxidized reactants — CO or \coo — involves more reaction intermediates and a change in the predominance of certain pathways. This explains the lower abundance of reduced organic compounds in an oxidized atmosphere, combined with the emergence of oxidation pathways that consume reaction intermediates, which have more influence for low C/O ratios. The addition of oxygen diversifies the chemistry, and in CO-rich and \coo-rich mixtures in particular, it leads to the formation of the oxidized organic compounds presented above. The combination of \chhhh and CO has a twofold effect. On the one hand, it provides carbon-rich atmospheres, which promote the efficient formation of hydrocarbons. On the other hand, it leads to moderately oxygenated atmospheres, which promote the formation of oxidized organic compounds without overly enhancing the pathways for hydrocarbon destruction.

Our study highlights the importance of adopting a synergistic approach combining experiments, modeling, and observations in order to deepen our understanding of exoplanet atmospheres. Each field offers unique perspectives, but through close collaboration their potential is greatly enhanced, and therefore strengthening these interdisciplinary efforts is essential.

\section{Perspectives}
The gas mixtures studied here cover a wide range of potential atmospheres. However, these are still fairly simple mixtures at a fixed temperature, which do not cover all possible atmospheric compositions of exoplanets. For example, the effect of temperature, other oxygen-containing reactants such as water in higher proportions, nitrogenous species, sulfur species, could be studied in the future. It would be very useful to study the detectability of photoproducts in detail, both for JWST observations and for future high-resolution ground-based observations.
%-----------------------------------------------------------------
\\
\\
\textit{Data availability.} Mass spectra, IR spectra, and 0D model data can be downloaded from Zenodo: \url{https://doi.org/10.5281/zenodo.17868179}.

\begin{acknowledgements}
This project has received funding from the European Research Council (ERC) under the ERC OxyPlanets projects (grant agreement No. 101053033). We would like to thank the anonymous referee for their excellent comments, which improved the presentation of our results. The authors thank Olivier Guaitella for kindly lending the multipass infrared cell, which was essential for this work. We would like to extend our warmest thanks to Joleen Csuka and Elisha Wesley Belibio for their valuable and helpful advice on proofreading and copyediting.
\end{acknowledgements}

\bibliographystyle{aa}
\bibliography{biblio}

% appendix
\begin{appendix}

\onecolumn

\section{Experimental identification of species}

\begin{figure}[ht!] 
  \begin{subfigure}{0.49\textwidth}
  \includegraphics[width=\textwidth]{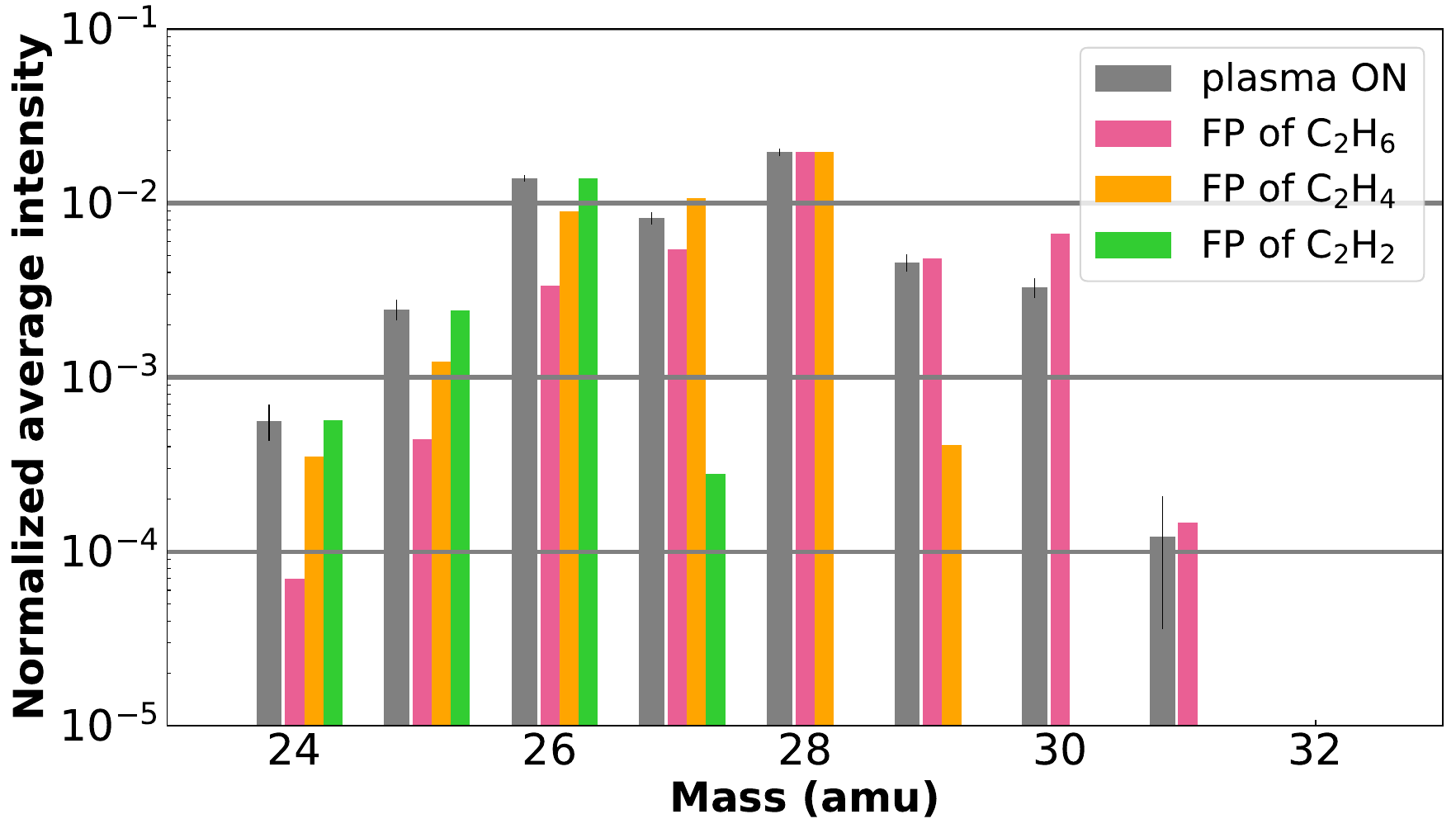}
  \caption{}
  \label{C2}
  \end{subfigure}
  \begin{subfigure}{0.49\textwidth}
  \includegraphics[width=\textwidth]{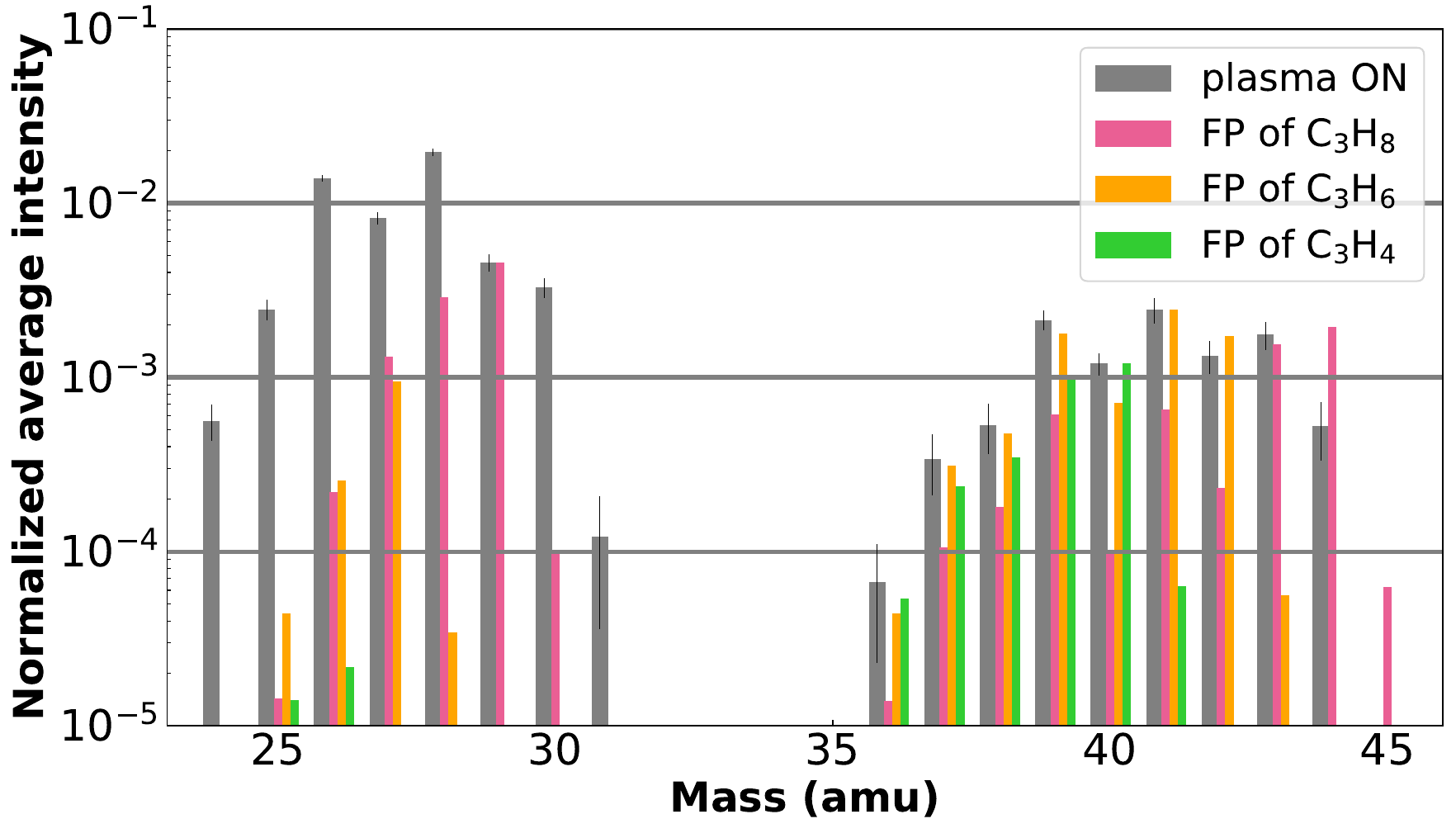}
  \caption{}
  \label{C3}
  \end{subfigure}
  \caption{Mass spectrum of an irradiated gas mixture of 95\% \hh + 5\% \chhhh (Gray). Fragmentation patterns (FP) of the two-carbon atoms (\ref{C2}) and three-carbon atoms (\ref{C3}) compounds (Pink for the alkanes, orange for the alkenes, green for the alkynes). Each FP is normalized to the intensity of the experimental spectrum at the m/z value corresponding to the most intense peak of the FP. 95\% uncertainty bars are displayed in black.}
  \label{FP_organics}
\end{figure}

\begin{figure}[ht!] 
  \centering
  \includegraphics[width=\textwidth]{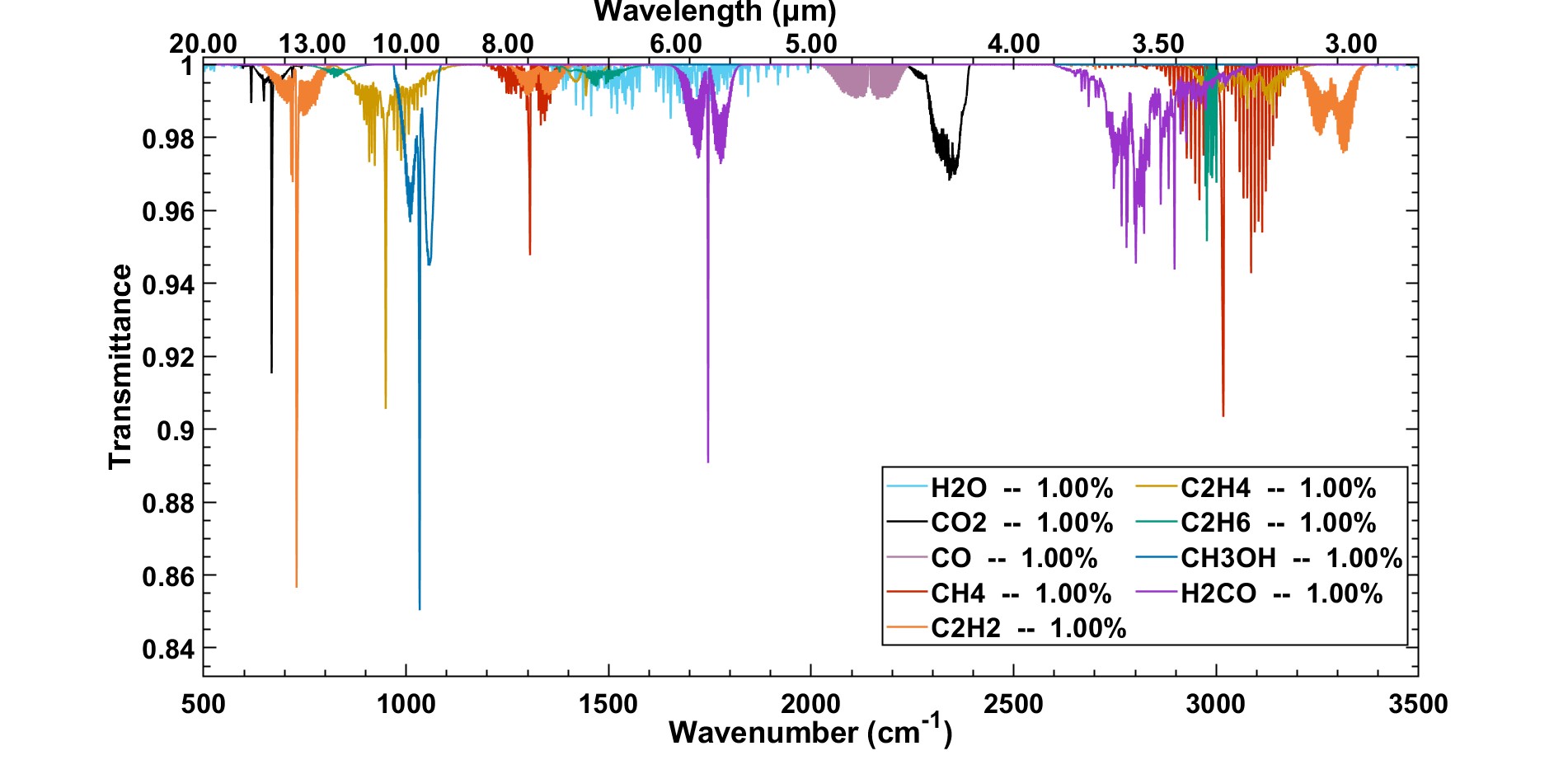}
  \caption{Infrared features calculated from the HITRAN database for molecules with up to 2 carbon atoms studied in this study, with a fixed composition of 1\% for each of them in a gas mixture of 1 mbar at 340 K.}
  \label{IR_ref_all}
\end{figure}

We report here the MS fragmentation patterns of hydrocarbon compounds with two and three atoms
(Figure \ref{FP_organics}), the reference IR spectra from the HITRAN database (Figure \ref{IR_ref_all}) and the main MS and IR features used to identity products (Table \ref{tab: features}).

\begin{table}[!htbp]
  \caption{Features of products identified by MS and IR spectroscopy.}
  \centering
  \begin{tabular}{c|c|c|c}
  \toprule
  \toprule
  \textbf{Molecules}    & \textbf{Main MS features (u)}   & \textbf{Main IR features (\cme)} & \textbf{Main IR features (µm)} \\ \midrule
  \cchh          & 26 - 25 - 24 - 27 - 28        & 730 [p] & 13.70 [p]            \\
  \cchhhh         & 28 - 27 - 26 - 25 - 24        & 950 [p] & 10.53 [p]            \\
  \cchhhhhh        & 28 - 27 - 30 - 29 - 26        & 2900->3000 [s] - 1300->1600 [s] & 3.45->3.33 [s] - 7.69->6.25 [s] \\
  \propdiene       & 40 - 39 - 38 - 37 - 36        & 1910->1990 [s] & 5.24->5.03 [s]         \\
  \propyne        & 40 - 39 - 38 - 37 - 36        & 633 [p] - 2100->2200 [s] & 15.80 [p] - 4.76->4.55 [s]    \\
  \ccchhhhhh       & 41 - 39 - 42 - 27 - 40        & 912 [p] & 10.96 [p]             \\
  \ccchhhhhhhh      & 29 - 28 - 27 - 44 - 43        & 2900->3000 [s] - 1300->1600 [s] & 3.45->3.33 [s] - 7.69->6.25 [s] \\
  \hhco          & 29 - 30 - 28             & 1745 [p] & 5.73 [p]            \\ 
  \chhhoh         & 31 - 32 - 29 - 15          & 1033 [p] & 9.68 [p]            \\  
  \chhhcho        & 29 - 44 - 43 - 15          & 1745 [s] & 5.73 [s]            \\ \midrule
  \multicolumn{4}{c}{[p] : single peak, narrow \; \; \; [s] : set of peaks, broad}
  \\ \bottomrule
  \end{tabular}
  %}
  \label{tab: features}
\end{table}

% \onecolumn

\end{appendix}
%  \bibliographystyle{aa} % style aa.bst
%  \bibliography{Yourfile} % your references Yourfile.bib

\end{document}